\documentclass[twocolumn]{aastex631}

\newcommand{\hb}{H$\beta$}

\newcommand{\oiii}{\textrm{[O\,\textsc{iii}]}}
\newcommand{\oii}{\textrm{[O\,\textsc{ii}]}}
\newcommand{\oi}{\textrm{[O\,\textsc{i}]}}
\newcommand{\nii}{\textrm{[N\,\textsc{ii}]}}
\newcommand{\neiii}{\textrm{[Ne\,\textsc{iii}]}}
\newcommand{\sii}{\textrm{[S\,\textsc{ii}]}}
\newcommand{\hii}{H \textsc{ii}}

\usepackage{tabularx}
\usepackage{amsmath}
\usepackage{amstext}
\usepackage{apjfonts}
\usepackage{graphicx}
\usepackage{tablefootnote}
\usepackage{amssymb}
\usepackage{latexsym}
\usepackage{longtable}
\usepackage{booktabs, amsmath, amssymb}

\received{October 9, 2026}

\submitjournal{ApJ}

\shorttitle{MAUNA Optical IFU}
\shortauthors{Kader et al.}
\graphicspath{{./}{figures/}}

\begin{document}

\title{Multiphase Analysis of (U)LIRG Nuclear Activity (MAUNA). I. Spatially Resolved Optical Emission-Line Diagnostics}

\correspondingauthor{Justin Kader}
\email{jkader@ipac.caltech.edu}

\author[0000-0002-6650-3757]{Justin A. Kader}
\affiliation{IPAC, California Institute of Technology, 1200 E. California Blvd., Pasadena, CA 91125, USA}

\author[0000-0002-1912-0024]{Vivian U}
\affiliation{IPAC, California Institute of Technology, 1200 E. California Blvd., Pasadena, CA 91125, USA}
\affiliation{Department of Physics and Astronomy, University of California, Irvine, 4129 Frederick Reines Hall, Irvine, CA 92697, USA}

\author[0000-0002-6570-9446]{Marina Bianchin}
\affiliation{Instituto de Astrof\'isica de Canarias, Calle V\'ia L\'actea, s/n, E-38205, La Laguna, Tenerife, Spain}
\affiliation{Departamento de Astrof\'isica, Universidad de La Laguna, E-38206, La Laguna, Tenerife, Spain}

\author[0000-0002-3139-3041]{Yiqing Song}
\affiliation{Max-Planck-Institut f\"{u}r Radioastronomie, Auf dem H\"{u}gel 69, 53121, Bonn, Germany}

\author[0000-0003-3474-1125]{George C. Privon}
\affiliation{National Radio Astronomy Observatory, 520 Edgemont Road, Charlottesville, VA 22903, USA}
\affiliation{Department of Astronomy, University of Virginia, 530 McCormick Road, Charlottesville, VA 22904, USA}
\affiliation{Department of Astronomy, University of Florida, P.O. Box 112055, Gainesville, FL 32611, USA}

\author[0000-0002-2596-8531]{Sabrina Stierwalt}
\affiliation{Occidental College, Physics Department, 1600 Campus Road, Los Angeles, CA 90042, USA}

\author[0000-0003-0699-6083]{Tanio Diaz-Santos}
\affiliation{Institute of Astrophysics, Foundation for Research and Technology-Hellas (FORTH), Heraklion, 70013, Greece}

\author[0000-0003-3498-2973]{Lee Armus}
\affiliation{IPAC, California Institute of Technology, 1200 E. California Blvd., Pasadena, CA 91125, USA}

\author[0000-0002-2688-1956]{Vassilis Charmandaris}
\affiliation{Institute of Astrophysics, Foundation for Research and Technology-Hellas (FORTH), Heraklion, 70013, Greece}
\affiliation{School of Sciences, European University Cyprus, Diogenes street, Engomi, 1516 Nicosia, Cyprus}
\affiliation{Department of Physics, University of Crete, Heraklion, 71003, Greece}

\author[0000-0001-7568-6412]{Ezequiel Treister}
\affiliation{Instituto de Alta Investigaci\'on, Universidad de Tarapac\'a: Arica, Arica y Parinacota Region, CL}

\author[0000-0003-2064-4105] {Rosalie McGurk}
\affiliation{W. M. Keck Observatory, 65-1120 Mamalahoa Hwy. Kamuela, HI 96743, USA}

\author[0000-0003-4268-0393]{Hanae Inami}
\affiliation{Hiroshima Astrophysical Science Center, Hiroshima University, 1-3-1 Kagamiyama, Higashi-Hiroshima, Hiroshima 739–8526, Japan}

\author[0000-0002-5807-5078]{Jeffrey A. Rich}
\affiliation{The Observatories of the Carnegie Institution for Science, 813 Santa Barbara Street, Pasadena, CA 91101, USA}

\author[0000-0002-1000-6081]{Sean T. Linden}
\affiliation{School of Physics and Astronomy, Rochester Institute of Technology, 84 Lomb Memorial Drive, Rochester, NY 14623, USA}
\affiliation{Steward Observatory, University of Arizona, 933 N Cherry Avenue, Tucson, AZ 85721, USA}

\author[0000-0001-7421-2944]{Anne M. Medling}
\affiliation{Department of Physics \& Astronomy and Ritter Astrophysical Research Center, University of Toledo, Toledo, OH 43606, USA}

\author[0000-0003-0057-8892]{Loreto Barcos-M\~{u}noz}
\affiliation{National Radio Astronomy Observatory, 520 Edgemont Road, Charlottesville, VA 22903, USA}
\affiliation{Department of Astronomy, University of Virginia, 530 McCormick Road, Charlottesville, VA 22904, USA}

\author[0000-0003-4693-6157]{Gabriela Canalizo}
\affiliation{Department of Physics and Astronomy, University of California, Riverside, 900 University Avenue, Riverside, CA 92521, USA}

\author[0000-0003-2638-1334]{Aaron S. Evans}
\affiliation{Department of Astronomy, University of Virginia, 530 McCormick Road, Charlottesville, VA 22904, USA}
\affiliation{National Radio Astronomy Observatory, 520 Edgemont Road, Charlottesville, VA 22903, USA}

\author[0000-0003-3917-6460]{Kirsten L. Larson}
\affiliation{AURA for the European Space Agency (ESA), Space Telescope Science Institute, 3700 San Martin Drive, Baltimore, MD 21218, USA}

\author[0000-0002-1158-6372]{Tianmu Gao}
\affiliation{Research School of Astronomy and Astrophysics, Australian National University, Weston Creek, ACT 2611, Australia}
\affiliation{ARC Centre of Excellence for All Sky Astrophysics in 3 Dimensions (ASTRO 3D), Canberra, ACT 2611, Australia}

\author[0000-0001-8490-6632]{Thomas Lai}
\affiliation{IPAC, California Institute of Technology, 1200 E. California Blvd., Pasadena, CA 91125, USA}

\author[0000-0003-4023-8657]{Laura Lenkic}\affiliation{IPAC, California Institute of Technology, 1200 E. California Blvd., Pasadena, CA 91125, USA}

\author[0009-0002-6248-3688]{Jaya Nagarajan-Swenson}
\affiliation{Department of Astronomy, University of Virginia, 530 McCormick Road, Charlottesville, VA 22904, USA}
\affiliation{National Radio Astronomy Observatory, 520 Edgemont Road, Charlottesville, VA 22903, USA}

\author[0000-0001-6028-8059]{Justin Howell}
\affiliation{IPAC, California Institute of Technology, 1200 E. California Blvd., Pasadena, CA 91125, USA}

\author[0000-0003-0532-6213]{Cristina M. Lofaro}
\affiliation{Department of Physics and Astronomy, University of Kentucky, 505 Rose Street, Lexington, KY 40506, USA}

\author[0000-0001-6919-1237]{Matthew A. Malkan}
\affiliation{Department of Physics \& Astronomy, University of California, Los Angeles, Los Angeles, CA 90095-1547, USA}

\author[0000-0003-4286-4475]{Mar\'ia S\'anchez-Garc\'ia}
\affil{Institute of Astrophysics, Foundation for Research and
Technology-Hellas (FORTH), Heraklion, 70013, Greece}

\author{Christopher D. Martin}
\affiliation{Cahill Center for Astronomy and Astrophysics, California Institute of Technology, 1200 E. California Blvd. MC 249-17, Pasadena, CA 91125, USA}

\author{Mateusz Matuszewski}
\affiliation{Cahill Center for Astronomy and Astrophysics, California Institute of Technology, 1200 E. California Blvd. MC 249-17, Pasadena, CA 91125, USA}

\author[0000-0003-0682-5436]{Claire E. Max}
\affiliation{Department of Astronomy \& Astrophysics, University of California, Santa Cruz, CA 95064, USA}

\author[0000-0002-8204-8619]{Joseph M. Mazzarella}
\affiliation{IPAC, California Institute of Technology, 1200 E. California Blvd., Pasadena, CA 91125, USA}

\author[0000-0002-0466-1119]{James D. Neill}
\affiliation{Cahill Center for Astronomy and Astrophysics, California Institute of Technology, 1200 E. California Blvd. MC 249-17, Pasadena, CA 91125, USA}

\author[0000-0001-5847-7934]{Nikolaus Z. Prusinski}
\affiliation{Cahill Center for Astronomy and Astrophysics, California Institute of Technology, 1200 E. California Blvd. MC 249-17, Pasadena, CA 91125, USA}

\author[0000-0002-0164-8795]{Raymond Remigio}
\affiliation{Department of Physics and Astronomy, University of California, Irvine, 4129 Frederick Reines Hall, Irvine, CA 92697, USA}

\author[0000-0002-1233-9998]{David Sanders}
\affiliation{Institute for Astronomy, University of Hawaii, 2680 Woodlawn Drive, Honolulu, HI 96822, USA}

\author[0000-0001-7291-0087]{Jason Surace}
\affiliation{IPAC, California Institute of Technology, 1200 E. California Blvd., Pasadena, CA 91125, USA}




\begin{abstract}
We present resolved optical integral-field spectroscopy of the 32 luminous and ultraluminous infrared galaxy systems of the Multiphase Analysis of (U)LIRG Nuclear Activity (MAUNA) sample, drawn from the GOALS survey and selected to have \emph{JWST}/MIRI medium-resolution spectrometer (MRS) observations. The data set combines new Keck/KCWI observations of 10 systems with archival VLT/MUSE cubes for 22 systems, covering H$\beta$ to \sii\ at $\sim1\arcsec$ spatial resolution ($\sim0.9$\,kpc at the sample median distance). Using the optical \nii/H$\alpha$ BPT diagram, we find that composite and LINER-like excitation together account for most of the classified area (median fractions $0.44$ and $0.26$), while Seyfert-like gas covers a median fraction of $0.02$ but carries a far larger proportion of the line luminosity where it is present. Both the projected size of the Seyfert-classified region, $R_{\rm Sy}=0.08$--$7.5$\,kpc, and its \oiii\ luminosity increase with the bolometric AGN luminosity, with $R_{\rm Sy}\propto L_{\rm AGN,bol}^{0.54\pm0.18}$, as expected for gas photoionized at a constant ionization parameter. The scatter about the luminosity relation depends on dust opacity: the \oiii\ luminosity per unit AGN luminosity falls by $1.09\pm0.25$ dex per unit depth of the $9.7$\,$\mu$m silicate absorption feature (Spearman $\rho=-0.60$, $p=0.001$). Optical excitation diagnostics therefore locate AGN-excited gas reliably but systematically underestimate the luminosity the AGN supplies to the host in the most obscured (U)LIRGs.
\end{abstract}



\section{Introduction} 
\label{sec:intro}

Starbursts and active galactic nuclei (AGN) inject radiative and kinetic energy into the surrounding interstellar medium (ISM). Their ionizing radiation controls the excitation of the gas across the host galaxy, and the galactic-scale winds they power regulate star formation and transport energy, momentum and metal-rich gas into the circumgalactic medium (CGM)~\cite[][]{Veilleux05,heckman17}. Ultra-luminous infrared galaxies (ULIRGs; $L_{\rm{IR}} \geq 10^{12} L_{\odot}$, and $\geq 10^{11} L_{\odot}$ for LIRGs) host the most extreme starburst and AGN activity in the local universe and are ideal laboratories for understanding how galactic evolution is impacted by AGN and stellar feedback~\cite[e.g.,][]{rodriguez-zaur13,pereira-santaella18}. Their AGN are also often deeply buried, so it remains an open question how much of the AGN's power reaches, and is visible in, the surrounding gas. Their outflows are a multiphase phenomenon, detected in ionized and neutral gas \cite[e.g.][]{Soto12,Leung21}, hot molecular gas~\cite[e.g.][]{dasyra16,emonts17}, warm molecular gas~\cite[e.g.][]{U13,U19}, and cold molecular gas~\cite[e.g.][]{GarciaBurillo15,barcos-munoz18,Izumi20}; recent IFU studies of GOALS systems include \cite{kader25}, \cite{kader26}, and \cite{song26}. Quantifying even their basic properties remains challenging~\citep{heckman17,Cicone18}, and characterizing them is a central aim of later MAUNA papers.

Because the influence of the AGN and starburst extends from the dust-shrouded nucleus to $\gtrsim$10\,kpc, where wide-field data also constrain outflow energetics and the gas deposited into the CGM~\citep{zhang18,Kakkad18,Cicone18}, a powerful approach is to combine high-resolution infrared integral-field unit (IFU) data that target the dust-shrouded nuclear region ($<$1 kpc) at 100 pc scales, which are themselves critical for understanding AGN feedback, with wide-field optical IFU data tracing the ionized gas from 1 kpc to scales of $>$10 kpc. One drawback of such observations is the limited field of view of the MIRI Medium Resolution Spectrometer (MRS) and the NIRSpec IFU ($\sim3\arcsec\times3\arcsec$ to $\sim7\arcsec\times7\arcsec$), making observations that cover the entire extent of feedback from nuclear to CGM scales very inefficient. Without coverage beyond the central kpc, neither the reach of the AGN's radiation nor the extent of its outflows can be measured. MAUNA addresses this limitation by pairing \emph{JWST} MIRI/MRS spectroscopy of the nuclei of 32 GOALS (U)LIRGs with wide-field optical integral-field data from Keck/KCWI and VLT/MUSE that map the same systems on 1--10\,kpc scales (Figure~\ref{fig:footprints}).

The primary scientific goals of the Multiphase Analysis of (U)LIRG Nuclear Activity (MAUNA) project are to understand whether outflows can effectively couple with and stir turbulence in the ISM, transport mass out of the galaxy, and drive metallicity gradients in the CGM. Characterization of ionized outflows using spatially resolved IFU observations is becoming increasingly common, and the full suite of optical diagnostic emission lines is an essential tool to understand the excitation mechanisms and ionization structure of the outflows. Unfortunately, capturing all lines needed for the analysis, from \oii~$\lambda\lambda$3726,3729 to \sii~$\lambda\lambda$6717,6731, is out of reach for many IFUs, whose sensitivity ends redward of the \oii\ doublet at the redshifts of local (U)LIRGs (e.g., MUSE, which starts at 4600--4750\,\AA). The Keck Cosmic Web Imager (KCWI), with its excellent blue sensitivity, wide wavelength coverage and large field of view (up to $35\arcsec\times24\arcsec$, or $\sim30\times20$\,kpc at the median distance of MAUNA), can measure all of these lines; its red arm, the Keck Cosmic Reionization Mapper (KCRM), extends the coverage to H$\alpha$ and \sii. The \oii~and \oiii~$\lambda$5007 lines can be used in conjunction to trace the full extent of galactic winds out to the extended low-ionization structure~\cite[][]{Collins09,morganti17,Chen19}. Later MAUNA papers will use this full suite of lines to characterize the density and ionization of the low-ionization gas, the ionized gas mass and the outflows; the present paper establishes the excitation structure on which those analyses build.

This first MAUNA paper addresses a narrower question: how much of the AGN power present in these systems do optical excitation diagnostics recover, and where? Mid-infrared diagnostics reveal AGN in many dusty galaxies whose optical spectra show no sign of one \citep{Goulding09}, and the optical classification of (U)LIRGs shifts from star-forming through composite to AGN-like along the merger sequence \citep{Yuan10}. Those studies used nuclear or integrated spectra. Here the excitation of every spaxel is mapped, and the luminosity of the AGN-excited gas is compared, system by system, with the AGN bolometric luminosity measured in the mid-infrared and with the depth of the 9.7\,$\mu$m silicate feature as a measure of nuclear obscuration.

The structure of this paper is as follows. In Section \ref{sec:sample} we describe how the MAUNA sample was selected and how it compares to the larger GOALS sample of (U)LIRGs from which it is drawn. In Section \ref{sec:observations_data} we describe the KCWI observations and data reduction, and provide a summary of the KCWI+MUSE data set as a whole. We present emission line flux maps, Baldwin-Phillips-Terlevich (BPT) emission line excitation diagnostic diagrams \citep{Baldwin81}, and galaxy maps color-coded by BPT classification in Section \ref{sec:analysis_results}. In Section \ref{sec:discussion} we discuss observed correlations between the derived sample-wide quantities, e.g., proportion of AGN-dominated regions compared to star formation-dominated regions in each galaxy, versus mid-infrared properties of the galaxies: the AGN fraction $f_{\rm{AGN}}$, the AGN bolometric luminosity and the depth of the 9.7\,$\mu$m silicate feature. Our summary and conclusions are presented in Section \ref{sec:summary_conclusions}, and Appendix~\ref{sec:app-robust} collects the tests of the robustness of the correlations.

\section{Sample}
\label{sec:sample}

The 32 systems that comprise MAUNA are a sub-sample of the 202 (U)LIRGs in the Great Observatories All-sky LIRG Survey (GOALS; \citealt{Armus09}). GOALS contains all of the 21 ULIRGs and 181 LIRGs within the \emph{IRAS} Revised Bright Galaxy Sample (RBGS; \citealt{sanders03}), which is a flux-limited complete sample of objects with $S_{\rm{60}} > 5.24$~Jy above a Galactic latitude of |$b$| > 5 degrees. MAUNA sources were selected based on the availability of both optical IFU observations and mid-infrared IFU observations with the \emph{James Webb Space Telescope} (\emph{JWST}) Mid-Infrared Instrument (MIRI) in its medium-resolution spectroscopy (MRS) mode; every MAUNA system has MIRI/MRS data. MAUNA contains all GOALS (U)LIRGs for which there were high quality \emph{JWST}/MIRI/MRS observations, from programs 1204, 1265, 1267, 1268, 1328, 1717, 1875, 1991, 3368 and 3869 (Section~\ref{sec:coverage}). Twenty-two MAUNA sources had archival VLT/MUSE optical IFU data available, 12 of which were observed additionally with Keck/KCWI to supplement the MUSE observations to cover the \oii~3726,29 \AA~doublet and the \neiii~3869 \AA~line. The remaining 10 MAUNA sources were observed with Keck/KCWI. For the 12 systems with both, the maps and measurements in this paper use the MUSE cube, which covers H$\beta$ to \sii\ in one exposure; their KCWI data will be used for \oii\ and \neiii\ in later papers. 

{\setlength{\tabcolsep}{7.42pt}
\begin{deluxetable*}{ll cc cccc c ccc}
\tabletypesize{\footnotesize}
\tablewidth{0pt}
\tablecaption{The MAUNA sample. Positions, luminosity distances, merger classes, silicate depths and AGN bolometric fractions are from the GOALS sample tables. $D_L$ is the luminosity distance \citep{Armus09}; the physical scale uses the angular-diameter distance, $D_A = D_L/(1+z)^2$. $L_{\rm IR}$ is the $8$--$1000$\,$\mu$m luminosity of the galaxy component(s) inside our IFU field, from the per-galaxy infrared luminosities of the GOALS sample, estimated from the \emph{Spitzer} 8, 24 and 70\,$\mu$m photometry of each component (\citealt{Stierwalt14}; S.~Stierwalt, private communication). The merger class follows \cite{Stierwalt13}: N = non-merger; a = pre-merger; b = early-stage merger; c = mid-stage merger; d = late-stage merger.  $-s_{9.7}$ is the apparent depth of the $9.7$\,$\mu$m silicate absorption feature from \cite{Stierwalt13}, positive for absorption and negative for emission.  $f_{\rm AGN}$ is the AGN contribution to the bolometric luminosity from \emph{Spitzer}/IRS mid-infrared diagnostics \citep{Diaz-Santos17}, and $L_{\rm AGN,bol} = 1.15 \sum_i f_{{\rm AGN},i}\, L_{{\rm IR},i}$ over the nuclei $i$ inside the IFU field, with $L_{\rm bol} = 1.15\, L_{\rm IR}$ \citep{Veilleux09}.
\label{tab:sample}}
\tablehead{
\colhead{IRAS ID} & \colhead{Alt.\ Name} &
\colhead{R.A.} & \colhead{Dec.} &
\colhead{$z$} & \colhead{$D_L$} & \colhead{Scale} & \colhead{$\log L_{\rm IR}$} &
\colhead{Merger} & \colhead{$-s_{9.7}$} & \colhead{$f_{\rm AGN}$} &
\colhead{$\log L_{\rm AGN,bol}$} \\
\colhead{} & \colhead{} &
\colhead{(J2000)} & \colhead{(J2000)} &
\colhead{} & \colhead{(Mpc)} & \colhead{(kpc/\arcsec)} & \colhead{($L_\odot$)} &
\colhead{class} & \colhead{} & \colhead{(bol.)} &
\colhead{($L_\odot$)}
}
\startdata
07251-0248 & \nodata & 07:27:37.5 & $-$02:54:55 & 0.0880 & 400 & 1.638 & 12.39 & d & 2.35 & 0.14 & 11.60 \\
09022-3615 & \nodata & 09:04:12.7 & $-$36:27:01 & 0.0600 & 271 & 1.169 & 12.31 & d & 0.88 & 0.19 & 11.65 \\
F01053-1746 & VV 114 & 01:07:47.2 & $-$17:30:25 & 0.0201 & 85.5 & 0.398 & 11.71 & c & 0.98 & 0.12 & 10.85 \\
F01364-1042 & \nodata & 01:38:52.9 & $-$10:27:11 & 0.0482 & 210 & 0.927 & 11.85 & d & 1.27 & 0.05 & 10.61 \\
F01417+1651 & III Zw 035 & 01:44:30.5 & +17:06:05 & 0.0274 & 119 & 0.547 & 11.64 & a & \nodata & 0.00 & \nodata \\
F04118-3207 & ESO 420-G013 & 04:13:49.7 & $-$32:00:25 & 0.0119 & 51.0 & 0.241 & 11.07 & N & 0.27 & 0.25 & 10.53 \\
F05189-2524 & \nodata & 05:21:01.4 & $-$25:21:45 & 0.0428 & 187 & 0.834 & 12.16 & d & 0.29 & 0.60 & 12.00 \\
F08572+3915 & \nodata & 09:00:25.4 & +39:03:54 & 0.0582 & 264 & 1.143 & 12.16 & d & 3.58 & 0.47 & 11.90 \\
F09111-1007 & \nodata & 09:13:38.8 & $-$10:19:20 & 0.0541 & 246 & 1.073 & 12.06 & b & 0.15 & 0.14 & 11.27 \\
F09320+6134 & UGC 05101 & 09:35:51.6 & +61:21:11 & 0.0394 & 177 & 0.794 & 12.02 & d & 0.78 & 0.25 & 11.47 \\
F10565+2448 & \nodata & 10:59:18.1 & +24:32:34 & 0.0431 & 197 & 0.878 & 12.08 & d & 0.75 & 0.04 & 10.75 \\
F12243-0036 & NGC 4418 & 12:26:54.6 & $-$00:52:39 & 0.0071 & 36.5 & 0.174 & 11.19 & N & 3.51 & 0.48 & 10.93 \\
F12540+5708 & Mrk 231 & 12:56:14.2 & +56:52:25 & 0.0422 & 192 & 0.857 & 12.57 & d & 0.48 & 0.53 & 12.36 \\
13120-5453 & \nodata & 13:15:06.3 & $-$55:09:23 & 0.0308 & 144 & 0.657 & 12.32 & d & 0.91 & 0.04 & 10.98 \\
F13126+2453 & IC 0860 & 13:15:03.5 & +24:37:08 & 0.0129 & 56.8 & 0.268 & 11.14 & N & 0.98 & 0.06 & 9.98 \\
F13428+5608 & Mrk 273 & 13:44:42.1 & +55:53:13 & 0.0373 & 173 & 0.779 & 12.21 & d & 1.37 & 0.31 & 11.76 \\
F14348-1447 & \nodata & 14:37:38.4 & $-$15:00:20 & 0.0823 & 387 & 1.602 & 12.39 & d & 1.36 & 0.09 & 11.40 \\
F14378-3651 & \nodata & 14:40:59.0 & $-$37:04:32 & 0.0676 & 315 & 1.340 & 12.23 & d & 1.14 & 0.05 & 10.99 \\
F14547+2449 & VV 340a & 14:57:00.8 & +24:37:04 & 0.0337 & 157 & 0.712 & 11.66 & b & 0.63 & 0.16 & 10.92 \\
F15250+3608 & \nodata & 15:26:59.4 & +35:58:38 & 0.0600 & 254 & 1.096 & 12.08 & d & 2.69 & 0.18 & 11.40 \\
F15327+2340 & Arp 220 & 15:34:57.2 & +23:30:11 & 0.0184 & 87.9 & 0.411 & 12.28 & d & 2.26 & 0.09 & 11.29 \\
F16504+0228 & NGC 6240 & 16:52:58.9 & +02:24:03 & 0.0243 & 116 & 0.536 & 11.93 & d & 0.92 & 0.11 & 11.03 \\
F17207-0014 & \nodata & 17:23:21.9 & $-$00:17:01 & 0.0428 & 198 & 0.883 & 12.46 & d & 1.26 & 0.05 & 11.22 \\
F19297-0406 & \nodata & 19:32:21.2 & $-$03:59:56 & 0.0857 & 395 & 1.625 & 12.45 & d & 1.05 & 0.05 & 11.21 \\
F19542+1110 & \nodata & 19:56:35.4 & +11:19:03 & 0.0660 & 295 & 1.259 & 12.12 & N & 0.74 & 0.08 & 11.09 \\
F20550+1655 & II Zw 096 & 20:57:23.9 & +17:07:39 & 0.0361 & 161 & 0.727 & 11.91 & c & 0.42 & 0.05 & 10.66 \\
F22491-1808 & \nodata & 22:51:49.2 & $-$17:52:23 & 0.0777 & 351 & 1.465 & 12.20 & d & 1.04 & 0.02 & 10.56 \\
F23007+0836 & NGC 7469 & 23:03:15.6 & +08:52:26 & 0.0163 & 70.8 & 0.332 & 11.58 & a & $-$0.06 & 0.24 & 11.02 \\
F23365+3604 & \nodata & 23:39:01.3 & +36:21:08 & 0.0645 & 287 & 1.219 & 12.20 & d & 1.41 & 0.05 & 10.96 \\
F10257-4339 & NGC 3256 & 10:27:51.3 & $-$43:54:13 & 0.0094 & 38.9 & 0.185 & 11.64 & d & 0.28 & 0.05 & 10.40 \\
F20551-4250 & ESO 286-IG019 & 20:58:26.8 & $-$42:39:00 & 0.0430 & 193 & 0.860 & 12.06 & d & 2.52 & 0.24 & 11.50 \\
F23128-5919 & ESO 148-IG002 & 23:15:46.7 & $-$59:03:16 & 0.0446 & 199 & 0.884 & 12.06 & c & 0.66 & 0.19 & 11.40 \\
\enddata
\tablecomments{Where the IFU field contains more than one GOALS nucleus with a measured AGN fraction, $L_{\rm IR}$ is their sum and $f_{\rm AGN}$ their $L_{\rm IR}$-weighted mean: IRAS F09111$-$1007 (W and E nuclei; $f_{\rm AGN}=0.13$, $0.18$; $\log L_{\rm IR}=11.95$, $11.39$) and II Zw 096 (W and E; $0.05$, $0.05$; $11.60$, $11.60$). For VV 340a only the northern galaxy, and for NGC 7469 only the primary (not IC 5283), lies in the field, so their $L_{\rm IR}$ refer to that galaxy alone. A nucleus counts as inside the field when BPT-classified spaxels lie within 2\arcsec\ of its GOALS position; IRAS F08572$+$3915 and F15250$+$3608, which have no BPT classification, take the GOALS nucleus nearest the pointing. III Zw 035 has $f_{\rm AGN}=0.00$ \citep{Diaz-Santos17}, so its $L_{\rm AGN,bol}$ is undefined.}
\end{deluxetable*}
}

MAUNA contains individual galaxies as well as systems at all stages of interaction from pre-merger to a fully coalesced, late-stage mergers \citep{Stierwalt13}. GOALS catalogs 40 galaxy nuclei within 90\arcsec\ of our 32 pointings, each with its own \emph{Spitzer}/IRS measurement \citep{Diaz-Santos17}; \cite{Diaz-Santos17} measure a non-zero mid-infrared AGN fraction for 34 of them, which is the sense in which we say that these nuclei host a mid-infrared AGN.
Table \ref{tab:sample} provides a summary of the basic properties of the MAUNA sample: IRAS ID, alternate name(s), coordinates, redshift, luminosity distance \citep{Armus09}, physical scale, infrared luminosity of the galaxies inside our field, merger class \citep{Stierwalt13}, silicate depth \citep{Stierwalt13}, AGN fraction \citep{Diaz-Santos17}, and AGN bolometric luminosity. The sample spans a redshift range of 0.007 < $z$~< 0.088, corresponding to a range in luminosity distances of 36.5 < $D_{\rm{L}}$~< 400 Mpc. The infrared luminosities of the galaxies in our fields range from 11.07 < log($L_{\rm{IR}}/L_\odot$) < 12.57. At the median scale of 0.86\,kpc\,arcsec$^{-1}$ (range 0.17--1.64), 1\arcsec\ corresponds to $\sim$0.9\,kpc. Among sources with a non-zero AGN fraction, $f_{\rm{AGN,bol}}$ ranges from 0.02 to 0.60, with AGN bolometric luminosities spanning 9.98 < log($L_{\rm{AGN,bol}}/L_\odot$) < 12.36.

The AGN bolometric luminosity is $L_{\rm AGN,bol}=1.15\sum_i f_{{\rm AGN},i}\,L_{{\rm IR},i}$, summed over the GOALS nuclei that lie inside our IFU field, each with its own AGN fraction and its own infrared luminosity (\citealt{Stierwalt14}; S.~Stierwalt, private communication). The choice of nuclei is geometric: a nucleus is included when BPT-classified spaxels lie within 2\arcsec\ of its position, regardless of where the Seyfert-classified gas lies, so that the optical quantities and $L_{\rm AGN,bol}$ refer to the same observed field. This affects four systems (Table~\ref{tab:sample}): IRAS F09111$-$1007 and II Zw 096, where both nuclei are in the field, and VV 340a and NGC 7469, where only one galaxy is. The AGN fractions of \cite{Diaz-Santos17} are derived from \emph{Spitzer}/IRS line and continuum ratios alone, calibrated following \cite{Veilleux09}; no infrared luminosity enters them, so pairing each nucleus's $f_{\rm AGN}$ with its own $L_{\rm IR}$ is the consistent choice. The IRS measurements integrate the nucleus within the Short-Low slit ($3\farcs6$ wide), a larger aperture than an IFU spaxel, but the mid-infrared emission of (U)LIRGs is compact \citep{Diaz-Santos10,Diaz-Santos11}, so we take $f_{\rm AGN}$ to describe the whole galaxy; extended star formation outside the slit would make $L_{\rm AGN,bol}$ an overestimate.

MAUNA samples the upper end of the infrared luminosity distribution of GOALS. Of the 32 systems in MAUNA, 21 are ULIRGs and 11 are LIRGs, compared to the 22 ULIRGs and 180 LIRGs in the parent GOALS sample. 
MAUNA therefore includes 21 of the 22 GOALS ULIRGs (95\%) but only 11 of its 180 LIRGs (6\%). IRAS F12112+0305 is the sole GOALS ULIRG absent from the sample due to not having been observed with \emph{JWST}/MIRI. The median infrared luminosity is log($L_{\rm{IR}}/L_\odot$) = 12.08, compared with 11.40 for GOALS. MAUNA also samples larger distances, with a median luminosity distance of 192.5 Mpc compared with 94.8 Mpc for the parent sample.

The sample spans all five merger classes tabulated for GOALS, including non-mergers, pre-mergers and early-, mid-, and late-stage mergers. However, their relative frequencies differ substantially from those in the parent population. Late-stage mergers account for 21 of the 32 MAUNA systems (66\%), compared with 54 of the 200 GOALS systems with an assigned merger class (27\%). The remaining MAUNA systems comprise four non-mergers, two pre-mergers, two early-stage mergers, and three mid-stage mergers. Thus, although MAUNA covers the range of interaction classes present in GOALS, its distribution is strongly weighted toward late-stage mergers, which largely reflects the luminosity selection, since most ULIRGs are late-stage mergers \citep{Stierwalt13}. 

The combination of wide interaction-stage coverage and nearly complete coverage of the GOALS ULIRGs provides an opportunity to examine resolved ionized gas excitation across the most infrared-luminous systems in the parent survey. The selected LIRGs extend this comparison to lower luminosities and include systems at earlier interaction stage. Interpretation of sample-wide trends must nevertheless account for the joint optical and mid-infrared observational selection and the uneven representation of luminosity and merger classes. In particular, flux distributions and excitation fractions measured for MAUNA characterize this observed sample, and any extrapolation to the full GOALS population requires accounting for these selection effects.

\section{Observations and Data}
\label{sec:observations_data}

\subsection{Keck/KCWI}
\label{sec:Data_kcwi}
Observations were conducted with KCWI \citep{morrisey18}, a wide-field integral field spectrograph optimized for high
throughput in the 3500–5600 \AA~bandpass on the Keck II telescope. KCWI was used for the MAUNA sources observable from Keck II that either had no optical IFU data, or had MUSE data that do not reach the \oii~3726,29\,\AA\ doublet, which traces the low-ionization gas expected to extend beyond the \oiii-emitting gas and will be studied in a forthcoming paper. The KCWI observations presented here were obtained on ten nights between 2020 August 19 and 2024 April 8 (UT) under nine Keck programs: 2020B\_U055 (PI: Max), 2021B\_U134 (PI: U), 2021B\_K318 (PI: Lewis),  2022A\_U213 (PI: U), 2022A\_K302 (PI: McGurk),  2023B\_K454 (PI: McGurk), 2023B\_U081 (PI: U), 2024A\_U305 (PI: U) and 2024A\_K453 (PI: McGurk). Observing conditions varied across the campaign, with nightly median seeing ranging from 0.$\arcsec$6~to 1.$\arcsec$4 FWHM, with a program median of $\sim$0.$\arcsec$9 ($\sim$0.8\,kpc at the sample median distance). Along the slices the KCWI data are therefore seeing-limited and oversampled by roughly a factor of three; across the slices the resolution is set by the 1$\farcs$358 slice width, which exceeds the seeing on all but the poorest nights, so in that direction the data are slice-width-limited rather than seeing-limited, and the KCWI point-spread function is anisotropic and broader than that of MUSE (the coadds are resampled onto 0$\farcs$291 square spaxels, as discussed below).

The observations were made with the instrument in several configurations, each defined by which dispersion grating, central wavelength, and slicer were used, and whether the red arm of the instrument (KCRM or KCWI-R) was used simultaneously with the blue arm (KCWI-B). In all cases, the large slicer was used, which yields a field of view (FoV) of 35.$\arcsec$3 $\times$~23.$\arcsec$6 ($\sim30\times20$\,kpc at the sample median distance) with slit width-limited angular resolution (0.$\arcsec$291 sampling rate along the 1$\farcs$358-wide slit). In conjunction with the large slicer, the low, medium, and high resolution blue gratings (BL, BM, and BH) have resolving powers of $R \sim$~900, 2000, and 4500, respectively. The central wavelength of the BL grating was tuned such that the instantaneous bandpass covers from the \oii~doublet to the \oiii~5007 \AA~ lines contiguously. When the BM grating was used, it was tuned such that either the \oii~doublet or the \oiii+H$\beta$ region was covered (both cannot be covered simultaneously with the instantaneous bandpass of 850 \AA. Each target observed with the BH grating (which has a small instantaneous bandpass of $\sim$400 \AA) was observed sequentially in BH1 (covering the \oii~doublet) and then in BH3 (covering the \oiii+H$\beta$ region). Since the commissioning of KCWI-R on 2023 June 4 (UT), all MAUNA sources observed with KCWI-B in BH configuration (16 systems; Table~\ref{tab:ifu-dataset}) were simultaneously observed with the high resolution red gratings, i.e., RH1 (tuned to cover the Na I D interstellar absorption line doublet at $\sim$5900 \AA), and RH2 (tuned to cover the H$\alpha$-\nii~complex and the \sii~6716,31 \AA~doublet). The configurations used to observe each KCWI target is listed in Table \ref{tab:ifu-dataset}. 

Integration times for individual BH exposures were between 1200 and 1320 s, and shorter for BL and BM (e.g., $\sim$650 s for VV 114 and $\sim$275 s for NGC 7469; Table~\ref{tab:ifu-dataset}). The exposure time was chosen in order to achieve a target SNR $\geq$~5 in the \neiii\ line observed at the detectable edge of the outflow in III Zw, one of the fainter objects in MAUNA making the exposure time estimate a conservative one. The red arm observations are more sensitive to cosmic rays, which are removed by median combining multiple (typically four) shorter 300 s exposures at the same position. The total exposure time in the red and blue high resolution gratings was approximately equal per-pointing. The detector was binned by a factor of 2 $\times$~2. Bias frames, dark frames, thorium-argon calibration lamp spectra, as well as dome flats and twilight flats were collected during each observing night.

Basic reduction and wavelength solution for the spectroscopic data were performed using the official KCWI data reduction pipeline \textsc{KCWI\_DRP} V1.3.1\footnote{https://kcwi-drp.readthedocs.io/en/latest/}. The reduction pipeline also performs flux calibration of the science frames using observations of standard stars, and applies sky subtraction using the dedicated sky pointings to account for contamination by sky emission lines and continuum. The instrumental contribution to the emission line widths was measured by fitting Gaussians to several emission lines with high signal-to-noise ratio (SNR) in the thorium-argon lamp exposures. 

The instrumental width (FWHM) was measured in each coadd from the residuals of unresolved night-sky lines (Section~\ref{sec:fits}): a Gaussian FWHM of 0.82--0.94\,\AA\ for BH3 and 1.18--1.37\,\AA\ for RH2, i.e.\ $\sigma_{\rm inst}\approx20$--26\,km\,s$^{-1}$ ($R\approx5000$--6500) at the fitted lines, and the nominal 5.0\,\AA\ ($\sigma_{\rm inst}\approx124$\,km\,s$^{-1}$, $R\approx1030$) for BL, where the sky lines are too blended to measure. The BH3 and RH2 arc-lamp widths are broader than the sky lines in the same cubes, so they cannot describe the coadds.

Most targets were mapped as small mosaics rather than single pointings. Of the 22 galaxies with reduced KCWI data, 21 were observed at two or more distinct pointings in at least one grating configuration; only IRAS F19542+1110 was covered by a single pointing due to time constraints during observing. Across the 85 (galaxy, night, arm, grating) configurations, 67 comprise multiple pointings, most commonly two, and up to nine for NGC 7469, eight for Arp 220 and six for VV 114. Adjacent pointings were deliberately overlapped: 93\% of the 228 pointing pairs share sky area, with mosaic steps of typically 18$\arcsec$ yielding an overlap of $\sim$33\% of a single IFU footprint; a smaller set of $\sim$3$\arcsec$ dither pairs share $\sim$88\% of their area. Each pointing covers 35.$\arcsec$3 $\times$~23.$\arcsec$6 on the large slicer with 2 $\times$~2 binning, sampled at 1.$\arcsec$358 $\times$~0.$\arcsec$291 per spaxel. Cubes were coadded with the \textsc{CWITools} package\footnote{https://cwitools.readthedocs.io/en/v0.8/} \citep{osullivan20}, strictly within a single (night, arm, grating, central wavelength) group, reprojected onto a common grid at the most frequent position angle of the contributing pointings and combined with full-pixel drizzle and propagated variance; dedicated sky pointings and any frames not sky-subtracted are excluded. The coadds are resampled onto square 0.$\arcsec$291 spaxels.

\subsection{VLT/MUSE}
\label{sec:Data_muse}
The MUSE observations are archival; the program and principal investigator for each dataset are given in Table~\ref{tab:muse-provenance}. All 22 datacubes were retrieved as ESO Phase 3 reduced products: 13 MUSE-DEEP combined cubes and 9 standard pipeline cubes, all taken in wide-field mode. A MUSE-DEEP product combines all the observing blocks of a program on one field into a single cube, whereas a standard product is the reduction of a single observing block, so the DEEP cubes reach fainter surface brightness (Appendix~\ref{sec:app-muse})\footnote{https://www.eso.org/sci/publications/announcements/sciann17032.html}. Each cube samples a $\sim1'\times1'$ field at $0\farcs2\times0\farcs2$ per spaxel. The delivered image quality recorded by the pipeline ranges from $0\farcs54$ to $1\farcs71$ FWHM with a median of $0\farcs82$, so the spatial resolution is seeing-limited and spatially oversampled by roughly a factor of four. The cubes span 4750–9350 \AA, or 4600–9350 \AA~for the subset taken with the extended blue cutoff, sampled at 1.25 \AA~per pixel with a band-averaged resolving power $R\simeq3000$, corresponding to a velocity FWHM of $\simeq100$ km s$^{-1}$.

\subsection{Multiwavelength coverage}
\label{sec:coverage}
\begin{figure*}
    \centering
    \includegraphics[width=1\linewidth]{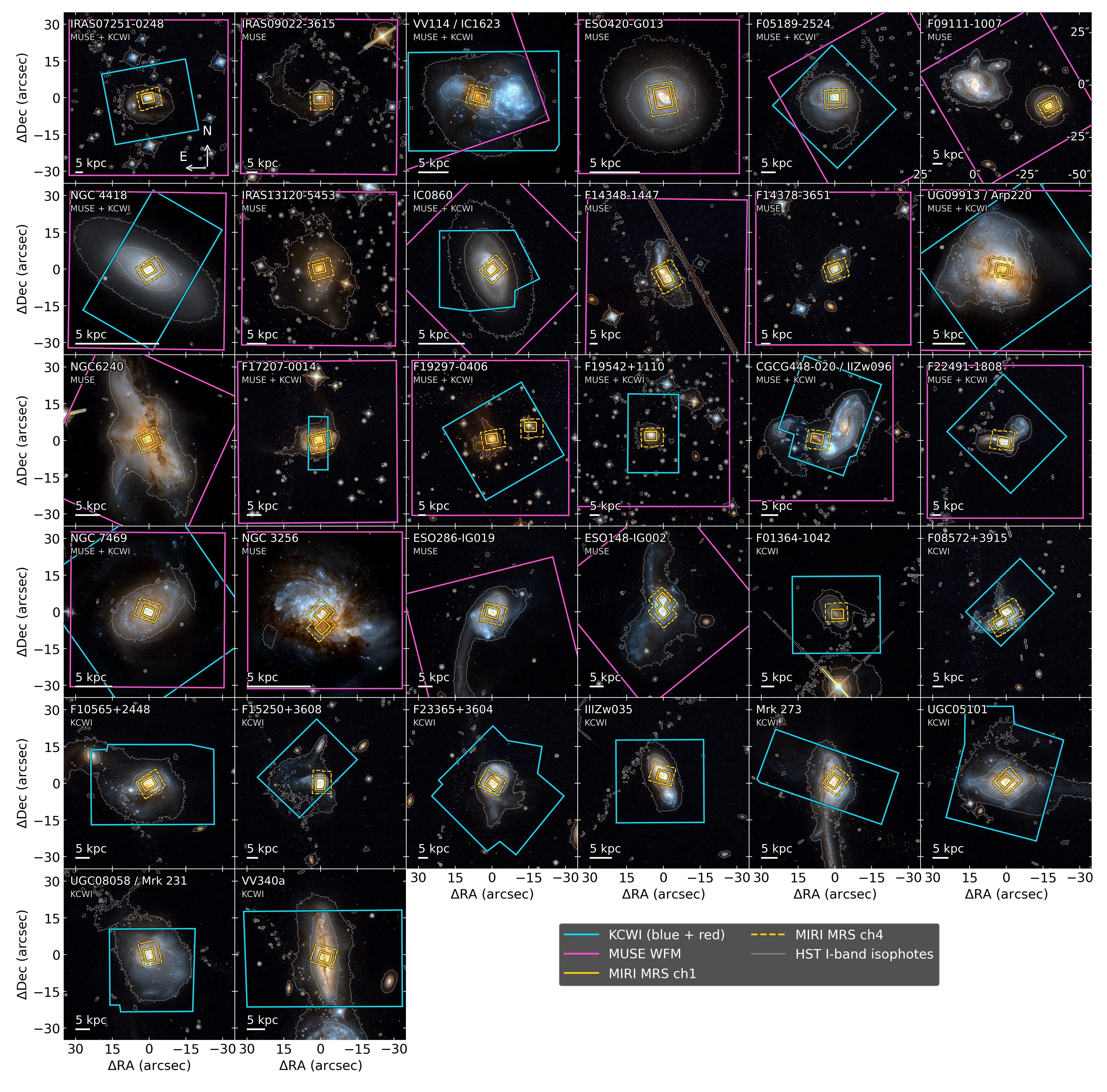}
    \caption{\emph{HST} views of the MAUNA sample with the footprints of the integral-field data. Each panel is a $70\arcsec\times70\arcsec$ color composite (red: $I$, F814W; green: mean of $I$ and $B$; blue: $B$, F435W or F438W, or F555W for NGC 4418) with an asinh stretch \citep{Lupton04}, centered on the position used for the maps of Figures~\ref{fig01:Halpha}--\ref{fig04:bptclass}, with north up and east left. The IRAS F09111$-$1007 panel spans $82\arcsec\times82\arcsec$, with its offsets marked inside it, to show both of its nuclei. Gray contours are the \emph{HST} $I$-band isophotes drawn on those maps. Outlines mark where each instrument has data: KCWI (cyan), the joint coverage of all reduced blue and red coadds, registered onto \emph{HST} as for the line maps; MUSE (magenta), the rotated rectangle enclosing the data of the cube used for the fits; and \emph{JWST} MIRI/MRS channel 1 (solid yellow; 4.9--5.7\,$\mu$m, the smallest field) and channel 4 (dashed yellow; the largest field). The panel labels list the optical IFU instruments; the bar in each panel is 5\,kpc, and the compass in the first panel applies to all.}
    \label{fig:footprints}
\end{figure*}

Figure~\ref{fig:footprints} shows each MAUNA system in \emph{HST} optical imaging together with the footprints of the KCWI, MUSE and \emph{JWST} MIRI/MRS data. The optical IFU data cover the galaxies on 1--10\,kpc scales, while the MRS fields (3--8\arcsec) are centered on the nuclei. The \emph{HST} images are the drizzled ACS/WFC F435W and F814W products of the GOALS \emph{HST} program (PID 10592) for 29 systems, and WFC3/UVIS F438W and F814W for ESO 420-G013 (PID 16914) and IC 0860 (PID 14715); NGC 4418 has no $B$-band image, and F555W (PID 14728) is used in its place. The MRS footprints are measured directly from the spaxels of the level-3 cubes in the Mikulski Archive for Space Telescopes (MAST), so they include every dither and mosaic tile at its true position angle. Dedicated MRS background pointings are not shown. The MRS observations of IRAS F09111$-$1007 target its western nucleus only. The MRS data come from \emph{JWST} programs 1204, 1265, 1267, 1268, 1328, 1717, 1875, 1991, 3368 and 3869; their reduction and the joint optical--mid-infrared analysis will be described in later MAUNA papers, for which this section and Figure~\ref{fig:footprints} serve as the reference.

\begin{deluxetable*}{ll @{\hspace{\dimexpr 2\tabcolsep + 14pt\relax}} cccc @{\hspace{\dimexpr 2\tabcolsep + 14pt\relax}} cccc @{\hspace{\dimexpr 2\tabcolsep + 14pt\relax}} cc}
\tabletypesize{\normalsize}
\tablewidth{0pt}
\tablecaption{Integral-field spectroscopy of the MAUNA sample.
$N$ is the number of individual on-source science exposures.  For KCWI--B each exposure yields one cube; for KCRM the pipeline stacks 3--4 consecutive exposures into each cube, and $N$ counts the underlying exposures rather than the stacks. $t$ is the total on-source shutter time.  $N_{\rm c}$ is the number of coadded cubes
delivered for that arm: the mosaicked pointings of one grating setting on one night combine into a single cube, so a source observed in two settings has $N_{\rm c}=2$, one per setting (the settings are listed under Gratings).
\label{tab:ifu-dataset}}
\tablehead{
\multicolumn{2}{c}{} & \multicolumn{4}{c}{KCWI--B} &
\multicolumn{4}{c}{KCWI--R} & \multicolumn{2}{c}{MUSE} \\
\cline{3-6} \cline{7-10} \cline{11-12}
\colhead{IRAS ID} & \colhead{Alternate Name} &
\colhead{$N$} & \colhead{$t$} & \colhead{Gratings} & \colhead{$N_{\rm c}$} &
\colhead{$N$} & \colhead{$t$} & \colhead{Gratings} & \colhead{$N_{\rm c}$} &
\colhead{Prod.} & \colhead{$t$} \\
\colhead{} & \colhead{} &
\colhead{} & \colhead{(hr)} & \colhead{} & \colhead{} &
\colhead{} & \colhead{(hr)} & \colhead{} & \colhead{} &
\colhead{} & \colhead{(hr)}
}
\startdata
07251-0248 & \nodata & 4 & 1.5 & BH1/BH3 & 2 & 16 & 1.3 & RH1/RH2 & 2 & DEEP & 2.5 \\
09022-3615 & \nodata & \nodata & \nodata & \nodata & \nodata & \nodata & \nodata & \nodata & \nodata & DEEP & 2.0 \\
F01053-1746 & VV 114 & 22 & 4.0 & BL/BM & 3 & \nodata & \nodata & \nodata & \nodata & DEEP & 0.8 \\
F01364-1042 & \nodata & 11 & 3.4 & BH1/BH3/BL & 3 & 29 & 2.4 & RH1/RH2 & 2 & \nodata & \nodata \\
F01417+1651 & III Zw 035 & 11 & 3.8 & BH1/BH3/BL & 3 & 16 & 1.3 & RH1/RH2 & 2 & \nodata & \nodata \\
F04118-3207 & ESO 420-G013 & \nodata & \nodata & \nodata & \nodata & \nodata & \nodata & \nodata & \nodata & DEEP & 0.6 \\
F05189-2524 & \nodata & 4 & 1.5 & BH1/BL & 2 & 16 & 1.3 & RH1/RH2 & 2 & DEEP & 2.0 \\
F08572+3915 & \nodata & 3 & 1.1 & BH1 & 1 & 12 & 1.0 & RH2 & 1 & \nodata & \nodata \\
F09111-1007 & \nodata & \nodata & \nodata & \nodata & \nodata & \nodata & \nodata & \nodata & \nodata & std & 0.7 \\
F09320+6134 & UGC 05101 & 12 & 2.6 & BH1/BL/BM & 3 & 24 & 2.0 & RH1/RH2 & 2 & \nodata & \nodata \\
F10565+2448 & \nodata & 10 & 2.9 & BH1/BL/BM & 3 & 8 & 0.7 & RH2 & 1 & \nodata & \nodata \\
F12243-0036 & NGC 4418 & 5 & 1.5 & BM & 1 & \nodata & \nodata & \nodata & \nodata & DEEP & 1.8 \\
F12540+5708 & Mrk 231 & 4 & 1.5 & BH1/BH3 & 2 & 16 & 1.3 & RH1/RH2 & 2 & \nodata & \nodata \\
13120-5453 & \nodata & \nodata & \nodata & \nodata & \nodata & \nodata & \nodata & \nodata & \nodata & std & 0.6 \\
F13126+2453 & IC 0860 & 6 & 1.8 & BH1/BH3/BM & 3 & 16 & 1.3 & RH1/RH2 & 2 & std & 0.8 \\
F13428+5608 & Mrk 273 & 4 & 1.4 & BH1/BH3 & 2 & 15 & 1.3 & RH1/RH2 & 2 & \nodata & \nodata \\
F14348-1447 & \nodata & \nodata & \nodata & \nodata & \nodata & \nodata & \nodata & \nodata & \nodata & DEEP & 1.3 \\
F14378-3651 & \nodata & \nodata & \nodata & \nodata & \nodata & \nodata & \nodata & \nodata & \nodata & std & 0.5 \\
F14547+2449 & VV 340a & 19 & 4.7 & BL/BM/BH3 & 4 & 28 & 2.3 & RH2 & 2 & \nodata & \nodata \\
F15250+3608 & \nodata & 1 & 0.4 & BH1 & 1 & 4 & 0.3 & RH2 & 1 & \nodata & \nodata \\
F15327+2340 & Arp 220 & 8 & 2.4 & BM & 1 & \nodata & \nodata & \nodata & \nodata & DEEP & 2.3 \\
F16504+0228 & NGC 6240 & \nodata & \nodata & \nodata & \nodata & \nodata & \nodata & \nodata & \nodata & DEEP & 1.3 \\
F17207-0014 & \nodata & 6 & 1.9 & BL & 1 & \nodata & \nodata & \nodata & \nodata & DEEP & 2.5 \\
F19297-0406 & \nodata & 5 & 1.5 & BH1/BH3 & 2 & 22 & 1.8 & RH1/RH2 & 2 & DEEP & 2.0 \\
F19542+1110 & \nodata & 4 & 1.1 & BH1/BH3 & 2 & 12 & 1.0 & RH1/RH2 & 2 & DEEP & 2.0 \\
F20550+1655 & II Zw 096 & 10 & 2.9 & BL/BM & 2 & \nodata & \nodata & \nodata & \nodata & DEEP & 0.9 \\
F22491-1808 & \nodata & 9 & 3.3 & BH1/BH3/BL & 4 & 36 & 3.0 & RH1/RH2 & 4 & std & 0.4 \\
F23007+0836 & NGC 7469 & 38 & 2.9 & BL/BM & 3 & \nodata & \nodata & \nodata & \nodata & std & 0.7 \\
F23365+3604 & \nodata & 11 & 3.9 & BH1/BH3/BL/BM & 5 & 28 & 2.3 & RH1/RH2 & 4 & \nodata & \nodata \\
F10257-4339 & NGC 3256 & \nodata & \nodata & \nodata & \nodata & \nodata & \nodata & \nodata & \nodata & std & 0.7 \\
F20551-4250 & ESO 286-IG019 & \nodata & \nodata & \nodata & \nodata & \nodata & \nodata & \nodata & \nodata & std & 0.9 \\
F23128-5919 & ESO 148-IG002 & \nodata & \nodata & \nodata & \nodata & \nodata & \nodata & \nodata & \nodata & std & 0.7 \\
\enddata
\tablecomments{Where an instrument or configuration was not used for a
source, the entry is \nodata.  MUSE products are the ESO reduced cubes: ``DEEP'' denotes a
MUSE-DEEP combined product, ``std'' a standard pipeline cube; their provenance
is given in Table~\ref{tab:muse-provenance}.}
\end{deluxetable*}

\section{Analysis and Results}
\label{sec:analysis_results}

Every MAUNA source has observations of the BPT lines \oiii~5007\,\AA, H$\beta$, \nii~6548,83\,\AA\ and H$\alpha$, and of \sii~6716,31\,\AA\ (used in the extended diagnostics of \citealt{Veilleux87}, which we defer to a later paper), with the exceptions of IRAS F08572+3915 and IRAS F15250+3608, whose only KCWI blue setting (BH1) covers \oii\ but neither H$\beta$ nor \oiii, so that neither can be placed on the BPT diagram. The incomplete coverage for these two systems is due to time and weather constraints during observing. IRAS F15250+3608 does have an H$\alpha$ map (Figure~\ref{fig01:Halpha}): its single RH2 pointing, one 300\,s exposure, was not coadded, so its map is drawn directly from the single cube, registered to \emph{HST} in the same way as the coadds. Spectral model fitting is used to produce spatially-resolved line surface brightness maps in each of the above lines, BPT diagrams, and resolved BPT classification maps of each galaxy. The spectral model fitting procedure and the results of the analysis are presented here.

\subsection{Spectral Fits}
\label{sec:fits}

Spectral models consisting of single-component Gaussian emission lines on a stellar continuum were fit to the KCWI and MUSE data cubes using \textsc{HyperCube} v0.5.1 \citep{Kader26_HyperCube}\footnote{https://github.com/jkader925/HyperCube}. \textsc{HyperCube} is a Python-based tool for fitting astronomical spectra (1D or IFU data), with both a graphical user interface which enables dynamic spectral model specification, fit inspection, and a fit rectification process, and a headless batch mode which takes a list of data cubes and a model template file at the command line. Both modes output a CSV file containing per-spaxel parameter initial guesses and best-fits. \textsc{HyperCube} uses \textsc{LMFIT}\footnote{https://github.com/lmfit/lmfit-py} (based on the Levenberg-Marquardt algorithm), which identifies best fit parameter values by non-linear least-squares minimization. Fits are done on a per-spaxel basis on two rest-frame windows: 4745-5155 \AA~(H$\beta$, \oiii~4959, 5007) and 6240-6830 \AA~(\oi~6300,64, \nii~6548,83, H$\alpha$, \sii~6716,31). MUSE covers both windows in one cube, while KCWI covers them in separate blue (BH3,BL) and red (RH2) cubes, fitted independently; the blue and red line measurements are then combined spaxel by spaxel by sky position (see below). Each fitted line is an intrinsic Gaussian broadened by the instrumental line-spread function (LSF), which is added in quadrature at the line's observed wavelength; the LSF is chosen per cube (instrument, grating and slicer). For MUSE we adopt the \cite{bacon17} sky-line fit, ${\rm FWHM}(\lambda)=5.866\times10^{-8}\lambda^2-9.187\times10^{-4}\lambda+6.040$\,\AA\ ($\lambda$ in \AA), so the resolving power rises from $R\approx1700$--1950 at H$\beta$ and \oiii\ to $R\approx2600$--2900 at H$\alpha$ over the redshifts of our sample. For KCWI the LSF is measured in each coadd from the residuals that sky subtraction leaves at unresolved night-sky lines (the OH lines for RH2 and [N\,\textsc{i}]~5199 for BH3), in the median spectrum of the faintest 30\% of spaxels. These residuals are single-signed and carry the instrument's own, flat-topped profile; a Gaussian fit gives FWHM~$=1.18$--1.37\,\AA\ for the nine RH2 coadds and 0.82--0.94\,\AA\ for the BH3 coadds (medians 1.27 and 0.84\,\AA, which are adopted for the two coadds with too few clean sky lines).
For BL, whose 1\,\AA\ sampling and blended sky lines prevent the measurement, we adopt the nominal 5.0\,\AA. These LSF values, not the arc-lamp widths, are the ones used in all fits. The BM, BH1 and RH1 data do not enter the fits. All line widths in subsequent MAUNA papers will be quoted as intrinsic velocity dispersions with the LSF removed in quadrature.

Each window's continuum is a \textsc{pPXF} \citep{Cappellari17} fit with E-MILES simple stellar populations \citep{Vazdekis16} (24 templates spanning 0.09--14 Gyr at a single, approximately solar metallicity, Kroupa initial mass function, and BaSTI isochrones, 2 kinematic moments, 8th-order multiplicative polynomial, and with no additive polynomial). A simple stellar population fit to the continuum is essential for H$\beta$ which often sits in significant stellar Balmer absorption and would be under-fit with a linear continuum model. Although E-MILES lacks very young stellar templates (which can generally bias continuum fits in ULIRGs), the Balmer absorption feature is still reliably fit because it is dominated by A stars with ages of hundreds of Myr, which are covered in the grid. Emission lines are masked during the stellar fit at each spaxel's own velocity. Spaxels with high continuum S/N are fitted on their own spectrum, while fainter ones are fitted on the mean of a 3$\times$3 to 11$\times$11 spaxel box, sized to reach a target continuum S/N of 40 per pixel, then rescaled to the spaxel's own line-free continuum level. 

Each emission line is fitted with a single Gaussian, fitted to the stellar-subtracted spectrum. \nii~6548 and 6583 velocity and width are tied to H$\alpha$, \oiii~4959 inherits velocity and width from \oiii~5007, and \sii~6731 inherits velocity and width from \sii~6716. For MUSE, H$\beta$ shares H$\alpha$ velocity and intrinsic width, accounting for LSF broadening. For KCWI, H$\beta$ is free since it comes from a different cube than H$\alpha$.
The flux ratios \nii~6583/6548 = 2.94 and \oiii~5007/4959 = 2.98 are fixed at their atomic values \citep{osterbrockferland06}, since both doublets arise from the same upper level. The density-sensitive \sii~6716/6731 ratio is free to vary between its non-degenerate diagnostic range (0.44-1.45). Intrinsic line dispersion is bounded to 10-650 km s$^{-1}$ so no line can collapse to zero-width. Line centroids are bounded to $\pm$500 km s$^{-1}$ around a per-spaxel velocity seed measured by a matched-filter cross-correlation of all lines simultaneously, which follows the rotation across each field.

A single Gaussian per line does not separate a broad-line-region or outflow component from the narrow line. In the type 1 nuclei of Mrk 231, NGC 7469 and IRAS F05189$-$2524, a single Gaussian fitted over broad and narrow H$\alpha$ raises the H$\alpha$ flux and the Balmer decrement and lowers \nii/H$\alpha$ in the nuclear spaxels, which moves them away from the Seyfert region of the BPT diagram; the Seyfert fractions of these systems are therefore lower limits, and their nuclear Balmer decrements are unreliable.
Outside these nuclei a single Gaussian per line is adequate for the line ratios used here, because the BPT ratios are formed from lines tied in velocity and width; multi-component kinematics are the subject of a later MAUNA paper.

A spaxel is fitted only where S/N $\geq$ 3 in H$\alpha$, H$\beta$, and \nii~6583 (typically $\sim$10-40\% of spaxels per cube). Every fitted line has an integrated flux, an uncertainty, and a S/N from $\sigma_{\rm{F}} = k\sigma_{\rm{n}}\sqrt{2\sqrt{\pi\sigma_{\rm{line}}\Delta\lambda}}$, with $\sigma_{\rm{n}}$~the robust scatter of the fit residual within $\pm$60 \AA. $k$~is a per-cube correlated-noise factor measured from fake lines injected at line-free positions in the same residuals ($\sim$1.0-1.2 for MUSE, $\sim$1.1-2.8 for KCWI, where drizzle-style resampling correlates adjacent pixels). Maps, BPT-class maps and BPT diagrams all share the same joint H$\alpha$-H$\beta$-\nii~S/N $\geq$~3 mask, with each line map masked on its own line. The classification uses the \cite{Kauffmann03} curve to distinguish ionized gas excited by star formation, the \cite{Kewley01} curve for the composite (star formation plus Seyfert or LINER) boundary, and \cite{schawinski07} to split LINERs from Seyferts \citep[see also][]{Kewley06}.  KCWI red and blue spaxels are paired by sky position through WCS-matching, after measuring the blue-red registration (shifts of 0-8$\arcsec$ across the sample); a red spaxel is used only where the blue spaxel containing it was also fitted.

\subsection{Emission Line Flux Maps}
\label{sec:linemaps}

\begin{figure*}
    \centering
    \includegraphics[width=1\linewidth]{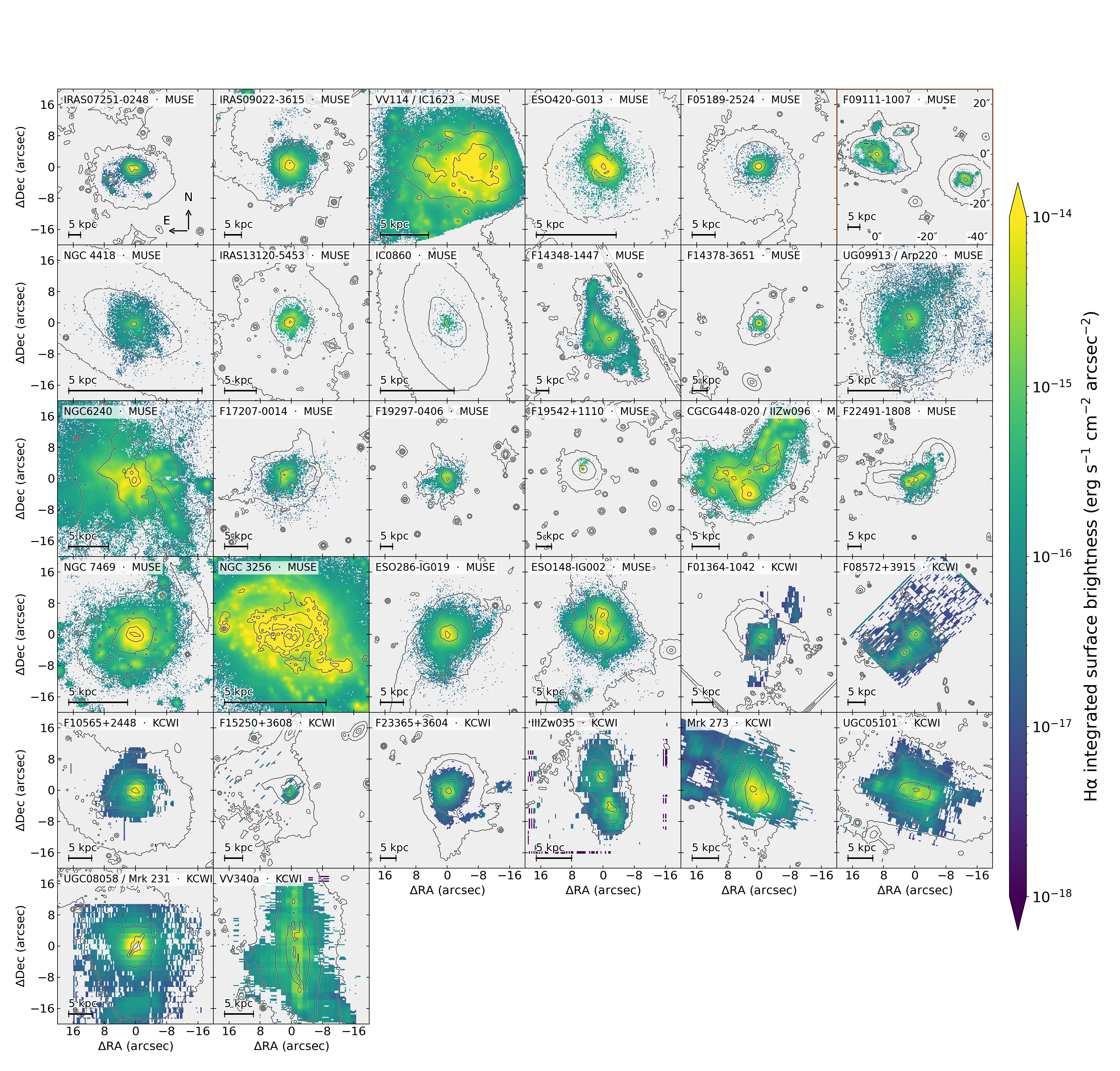}
    \caption{H$\alpha$ surface brightness across the MAUNA sample: all 32 systems, 22 observed with MUSE and 10 with KCWI.  Each panel spans $40\arcsec\times40\arcsec$ centered on the catalog position of the source (for IRAS F19297$-$0406, on its H$\alpha$ nucleus, 17\arcsec\ from the catalog position; Section~\ref{sec:linemaps}), with north up and east left, and is labeled with the source name and the instrument. The IRAS F09111$-$1007 panel instead spans $62\arcsec\times62\arcsec$, so that it includes the western nucleus $36\arcsec$ from the eastern one; its offsets are marked inside that panel. A spaxel is shown where its fitted H$\alpha$ line has S/N~$\geq3$; no fit-quality cut is applied.  Fluxes are divided by the spaxel solid angle, so that the MUSE and KCWI panels are directly comparable and can share the single logarithmic color scale at right, which runs from $10^{-18}$ to $10^{-14}$\,erg\,s$^{-1}$\,cm$^{-2}$\,arcsec$^{-2}$ in every panel.  The bar in each panel is 5\,kpc at the distance of that source, and the compass in the first panel applies to all.  Gray contours are \emph{HST} F814W isophotes, four logarithmically spaced levels beginning $5\sigma$ above the background after $0\farcs2$ smoothing, for all 32 cubes, each registered to the \emph{HST} imaging. Figure~\ref{fig:footprints} shows a wider field with the instrument footprints.  The map of IRAS F15250$+$3608 comes from a single 300\,s RH2 exposure that was not coadded; the isolated spaxels scattered away from its nucleus are noise at the S/N~$\geq3$ threshold.}
    \label{fig01:Halpha}
\end{figure*}

\begin{figure*}
    \centering
    \includegraphics[width=1\linewidth]{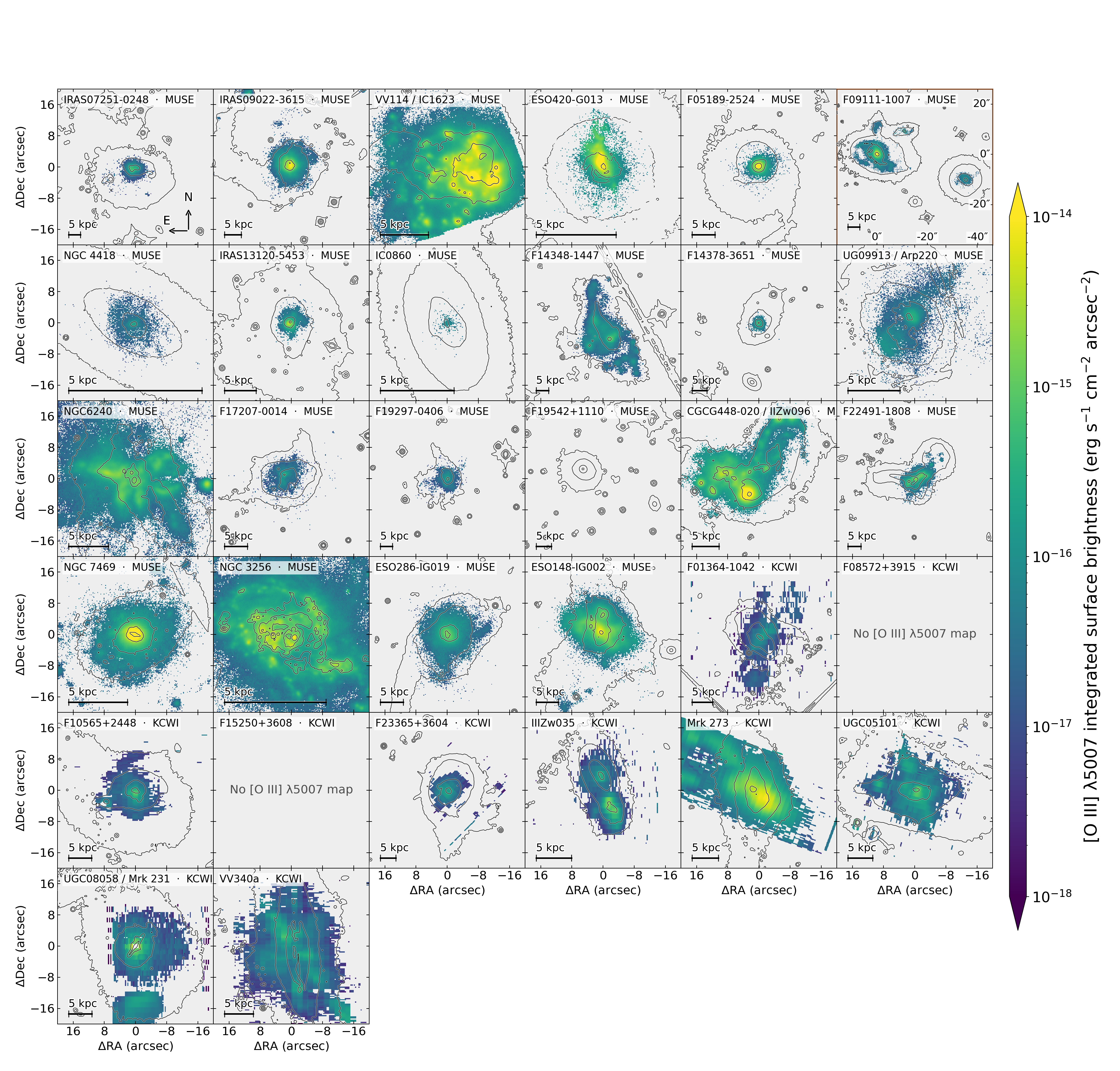}
    \caption{As Figure \ref{fig01:Halpha}, but for \oiii~$\lambda$5007: 30 of the 32 systems, 21 with MUSE and 8 with KCWI.  No map is shown for IRAS F08572$+$3915 and F15250$+$3608, whose only KCWI blue setting (BH1) covers neither \hb\ nor \oiii; for F08572$+$3915, for example, \oiii~$\lambda$5007 falls at 5298\,\AA, outside its 3858--4234\,\AA\ coverage.}
    \label{fig02:Oiii}
\end{figure*}

Emission line flux maps for the MAUNA sample are presented in Figures \ref{fig01:Halpha} and \ref{fig02:Oiii}; each line map shows the spaxels where that line has S/N $\geq$~3. The BPT diagrams, the class maps and Table~\ref{tab:derived} use the joint mask, S/N $\geq$~3 in H$\alpha$, H$\beta$ and \nii~6583, which leaves $N_{\rm{spax}}$ spaxels per source (column 4 of Table \ref{tab:derived}). Each map is centered on the catalog position of Table~\ref{tab:sample}; for multiple systems this is the position of the targeted component. The MUSE mosaic of IRAS F09111$-$1007 covers both of its galaxies; its catalog position coincides with the eastern nucleus (to within 1\arcsec), and its panels are widened to $62\arcsec\times62\arcsec$ to include the western nucleus, 36\arcsec\ away. IRAS F19297$-$0406 is a different case: its GOALS catalog position (Table~\ref{tab:sample}) lies on blank sky 17\arcsec\ west-northwest of the galaxy, so its panels are centered instead on the H$\alpha$ nucleus we measure, 19$^{\rm h}$32$^{\rm m}$22\fs3, $-04^\circ00\arcmin02\arcsec$, which agrees to within 2\arcsec\ with the \emph{Spitzer}/IRS positions of \cite{Stierwalt13} and \cite{Inami13}. The physical area covered by the spaxels that remain in the joint mask ($A_{\rm{BPT}}$ in Table \ref{tab:derived}) and spans 0.6--514 kpc$^2$ (mean 145 kpc$^2$), which is set in part by the intrinsic galaxy size but also by its distance and the field of view (Figure~\ref{fig:footprints}).

The spectral models produce per-spaxel line fluxes in erg s$^{-1}$~cm$^{-2}$. This quantity carries a factor $A_{\rm{spax}}$, the solid angle of the instrument spaxel, meaning that for a shared flux scale (colorbar), a KCWI spaxel (resampled to 0$\farcs$291 square, 0.085\,arcsec$^2$) would look about twice as bright as a MUSE spaxel (0.04\,arcsec$^2$) of the same intrinsic surface brightness. Dividing the emission line flux by the spaxel solid angle (surface brightness, erg s$^{-1}$~cm$^{-2}$~arcsec$^{-2}$) ensures a shared physical scaling across the sample regardless of instrument.

The sample spans a range in ionized gas morphology, from compact nuclear sources (e.g., IRAS 07251-0248, IRAS F05189-2524, IRAS F13120-5453, IC 0860, IRAS F19542+1110) to sources with ionized gas structure that clearly tracks extended star-forming regions (e.g., VV 114, II Zw 096, NGC 7469, NGC 3256), to cases where very extended ionized gas exists without an obvious broadband optical (star-forming) component (e.g., IRAS F01364-1042, III Zw 035, Mrk 273, UGC 05101, Mrk 231, VV 340a). Sources with two nuclei in the optical line maps include ESO148-IG002, IRAS F08572+3915, III Zw 035, and Mrk 273. Several systems with two nuclei at other wavelengths show only one in the line maps, either because one nucleus is obscured or because the pair is unresolved at our resolution (e.g., Arp 220, IRAS F14348$-$1447, IRAS 07251$-$0248, NGC 3256). IRAS F19542+1110 was covered by a single compact pointing and only 10 of its spaxels pass the mask, which is why it has so few spaxels. 

The H$\alpha$ luminosity summed over the remaining spaxels, $\log L(H\alpha)$, is listed in column (13) of Table \ref{tab:derived}.  It spans 38.92 $<$ $\log L(H\alpha)/{\rm erg\,s^{-1}}$ $<$ 42.39 with a median of 41.69, the extremes being IRAS F19542+1110, for which only 10 spaxels pass the mask, and II Zw 096 and VV 114, whose star-forming complexes are bright and extended. This quantity is observed rather than intrinsic: no correction for extinction has been applied, and in systems with $A_V$ of several magnitudes across the line-emitting gas (Appendix~\ref{sec:app-robust}) the intrinsic luminosity is higher by a correspondingly large factor. The BPT ratios, which pair lines close in wavelength, are almost entirely immune to this extinction.  It is also summed only over spaxels that pass the joint S/N mask, so gas below that threshold contributes nothing, and a single Gaussian per line omits any broad or outflowing component that it does not capture (Section~\ref{sec:fits}).  These effects act in the same direction, and $\log L(H\alpha)$ should therefore be read as a lower bound on the H$\alpha$ luminosity of the ionized gas in each system, and not as a total galaxy H$\alpha$ luminosity.

\begingroup
\setlength{\tabcolsep}{3.5pt}
{\setlength{\tabcolsep}{4.27pt}
\begin{deluxetable*}{ll c rr cccc cc cc}
\tabletypesize{\normalsize}
\tablewidth{0pt}
\tablecaption{Quantities derived from the MAUNA optical IFU fits.  All columns
are measured over the spaxels that pass a joint signal-to-noise cut of S/N~$\geq3$ in H$\alpha$, \hb\ and \nii~6583 (column 4), which is the same mask used for the BPT diagrams and the BPT classification maps.  Spaxels in which \oiii~5007 is undetected enter as $3\sigma$ upper limits: they are classified when the limit alone places them below the \cite{Kauffmann03} star-forming boundary, and are otherwise undetermined and excluded from columns 6--11 (a median of 7\% of the mask per system, at most 80\%).  Columns 6--9 are fractions of the classified spaxels in each excitation class, so they sum to unity; columns 10 and 11 are the Seyfert fraction weighted by H$\alpha$ and by \oiii~5007 luminosity instead of by area. $R_{\rm Sy} \equiv (A_{\rm Sy}/\pi)^{1/2}$ is the equivalent radius of the Seyfert-classified area, i.e.\ the radius of a circle of the same area, and is quoted whether or not that area is contiguous.  H$\alpha$ is summed over the mask and is not corrected for extinction.
\label{tab:derived}}
\tablehead{
\colhead{IRAS ID} & \colhead{Alt.\ Name} & \colhead{Instr.} &
\colhead{$N_{\rm spax}$} & \colhead{$A_{\rm BPT}$} &
\colhead{$f_{\rm SF}$} & \colhead{$f_{\rm comp}$} & \colhead{$f_{\rm LIN}$} &
\colhead{$f_{\rm Sy}$} & \colhead{$f_{\rm Sy}(H\alpha)$} &
\colhead{$f_{\rm Sy}$(\oiii)} & \colhead{$R_{\rm Sy}$} &
\colhead{$\log L(H\alpha)$} \\
\colhead{} & \colhead{} & \colhead{} & \colhead{} & \colhead{(kpc$^{2}$)} &
\colhead{} & \colhead{} & \colhead{} & \colhead{} & \colhead{} & \colhead{} &
\colhead{(kpc)} & \colhead{(erg\,s$^{-1}$)} \\
\colhead{(1)} & \colhead{(2)} & \colhead{(3)} & \colhead{(4)} & \colhead{(5)} &
\colhead{(6)} & \colhead{(7)} & \colhead{(8)} & \colhead{(9)} & \colhead{(10)} &
\colhead{(11)} & \colhead{(12)} & \colhead{(13)}
}
\startdata
07251-0248 & \nodata & MUSE & 1\,285 & 137.9 & 0.57 & 0.41 & 0.01 & 0.01 & 0.00 & 0.01 & 0.69 & 41.85 \\
09022-3615 & \nodata & MUSE & 3\,319 & 181.4 & 0.23 & 0.70 & 0.04 & 0.03 & 0.00 & 0.01 & 1.33 & 42.20 \\
F01053-1746 & VV 114 & MUSE & 27\,903 & 176.8 & 0.56 & 0.43 & 0.00 & 0.01 & 0.00 & 0.00 & 0.56 & 42.37 \\
F01364-1042 & \nodata & KCWI & 1\,254 & 91.5 & 0.04 & 0.20 & 0.75 & 0.01 & 0.00 & 0.00 & 0.55 & 40.74 \\
F01417+1651 & III Zw 035 & KCWI & 2\,209 & 56.1 & 0.36 & 0.36 & 0.27 & 0.01 & 0.00 & 0.00 & 0.32 & 40.95 \\
F04118-3207 & ESO 420-G013 & MUSE & 3\,254 & 7.6 & 0.05 & 0.24 & 0.40 & 0.31 & 0.34 & 0.76 & 0.86 & 41.22 \\
F05189-2524 & \nodata & MUSE & 925 & 25.7 & 0.00 & 0.02 & 0.63 & 0.35 & 0.87 & 0.90 & 1.64 & 41.35 \\
F08572+3915 & \nodata & KCWI & \nodata & \nodata & \nodata & \nodata & \nodata & \nodata & \nodata & \nodata & \nodata & \nodata \\
F09111-1007 & \nodata & MUSE & 5\,462 & 251.5 & 0.59 & 0.35 & 0.02 & 0.04 & 0.31 & 0.67 & 1.69 & 42.02 \\
F09320+6134 & UGC 05101 & KCWI & 3\,252 & 174.2 & 0.11 & 0.27 & 0.47 & 0.16 & 0.04 & 0.18 & 2.93 & 41.50 \\
F10565+2448 & \nodata & KCWI & 1\,794 & 117.5 & 0.05 & 0.59 & 0.34 & 0.03 & 0.00 & 0.02 & 0.97 & 41.56 \\
F12243-0036 & NGC 4418 & MUSE & 3\,734 & 4.5 & 0.02 & 0.86 & 0.11 & 0.01 & 0.00 & 0.01 & 0.10 & 39.49 \\
F12540+5708 & Mrk 231 & KCWI & 4\,504 & 281.0 & 0.15 & 0.47 & 0.29 & 0.10 & 0.02 & 0.09 & 2.91 & 42.27 \\
13120-5453 & \nodata & MUSE & 1\,201 & 20.7 & 0.01 & 0.56 & 0.36 & 0.07 & 0.15 & 0.16 & 0.64 & 41.25 \\
F13126+2453 & IC 0860 & MUSE & 201 & 0.6 & 0.00 & 0.01 & 0.99 & 0.00 & 0.00 & 0.00 & $<$0.03 & 38.96 \\
F13428+5608 & Mrk 273 & KCWI & 6\,889 & 355.1 & 0.04 & 0.18 & 0.28 & 0.50 & 0.83 & 0.88 & 7.48 & 41.94 \\
F14348-1447 & \nodata & MUSE & 5\,009 & 514.2 & 0.46 & 0.51 & 0.03 & 0.00 & 0.00 & 0.00 & 0.85 & 42.08 \\
F14378-3651 & \nodata & MUSE & 366 & 26.3 & 0.09 & 0.55 & 0.36 & 0.01 & 0.02 & 0.02 & 0.21 & 41.19 \\
F14547+2449 & VV 340a & KCWI & 6\,628 & 285.4 & 0.11 & 0.30 & 0.48 & 0.11 & 0.07 & 0.29 & 3.18 & 41.79 \\
F15250+3608 & \nodata & KCWI & \nodata & \nodata & \nodata & \nodata & \nodata & \nodata & \nodata & \nodata & \nodata & \nodata \\
F15327+2340 & Arp 220 & MUSE & 8\,502 & 57.4 & 0.00 & 0.10 & 0.90 & 0.00 & 0.00 & 0.00 & 0.11 & 40.73 \\
F16504+0228 & NGC 6240 & MUSE & 35\,978 & 413.5 & 0.05 & 0.45 & 0.47 & 0.02 & 0.01 & 0.03 & 1.67 & 42.08 \\
F17207-0014 & \nodata & MUSE & 1\,583 & 49.4 & 0.09 & 0.64 & 0.26 & 0.02 & 0.01 & 0.02 & 0.50 & 41.18 \\
F19297-0406 & \nodata & MUSE & 1\,010 & 106.7 & 0.07 & 0.64 & 0.27 & 0.02 & 0.01 & 0.02 & 0.82 & 41.59 \\
F19542+1110 & \nodata & MUSE & 10 & 0.6 & 1.00 & 0.00 & 0.00 & 0.00 & 0.00 & \nodata & $<$0.14 & 38.92 \\
F20550+1655 & II Zw 096 & MUSE & 10\,717 & 226.6 & 0.88 & 0.12 & 0.00 & 0.00 & 0.00 & 0.00 & 0.08 & 42.39 \\
F22491-1808 & \nodata & MUSE & 1\,683 & 144.5 & 0.57 & 0.43 & 0.01 & 0.00 & 0.00 & 0.00 & 0.17 & 41.94 \\
F23007+0836 & NGC 7469 & MUSE & 18\,780 & 82.8 & 0.43 & 0.49 & 0.05 & 0.03 & 0.22 & 0.37 & 0.85 & 42.00 \\
F23365+3604 & \nodata & KCWI & 809 & 102.1 & 0.11 & 0.77 & 0.09 & 0.03 & 0.01 & 0.02 & 0.92 & 41.39 \\
F10257-4339 & NGC 3256 & MUSE & 56\,660 & 77.6 & 0.44 & 0.53 & 0.03 & 0.01 & 0.00 & 0.00 & 0.36 & 42.07 \\
F20551-4250 & ESO 286-IG019 & MUSE & 5\,912 & 174.9 & 0.16 & 0.75 & 0.09 & 0.01 & 0.00 & 0.00 & 0.61 & 41.79 \\
F23128-5919 & ESO 148-IG002 & MUSE & 6\,486 & 202.7 & 0.19 & 0.61 & 0.00 & 0.20 & 0.16 & 0.21 & 3.57 & 42.18 \\
\enddata
\tablecomments{Rows are in the order of Table~\ref{tab:sample}.  IRAS
F08572$+$3915 and F15250$+$3608 have no BPT classification, because their only KCWI blue setting (BH1) covers neither \hb\ nor \oiii, and are listed for completeness. A limit on $R_{\rm Sy}$ is the equivalent radius of a single spaxel, quoted for the systems with no Seyfert-classified spaxel.  $R_{\rm Sy}$ is measured in projection, which makes it a lower bound on the physical extent of the AGN-ionized gas, but it is not corrected for the point-spread function, which for a region smaller than the PSF overestimates its area; for marginally resolved systems the two effects act in opposite directions.  Column 11 is undefined for IRAS F19542$+$1110, whose classified spaxels all have \oiii\ upper limits only.}
\end{deluxetable*}
}
\endgroup

\subsection{Excitation Diagrams and Maps}
\label{sec:BPT}

\begin{figure*}
    \centering
    \includegraphics[width=1\linewidth]{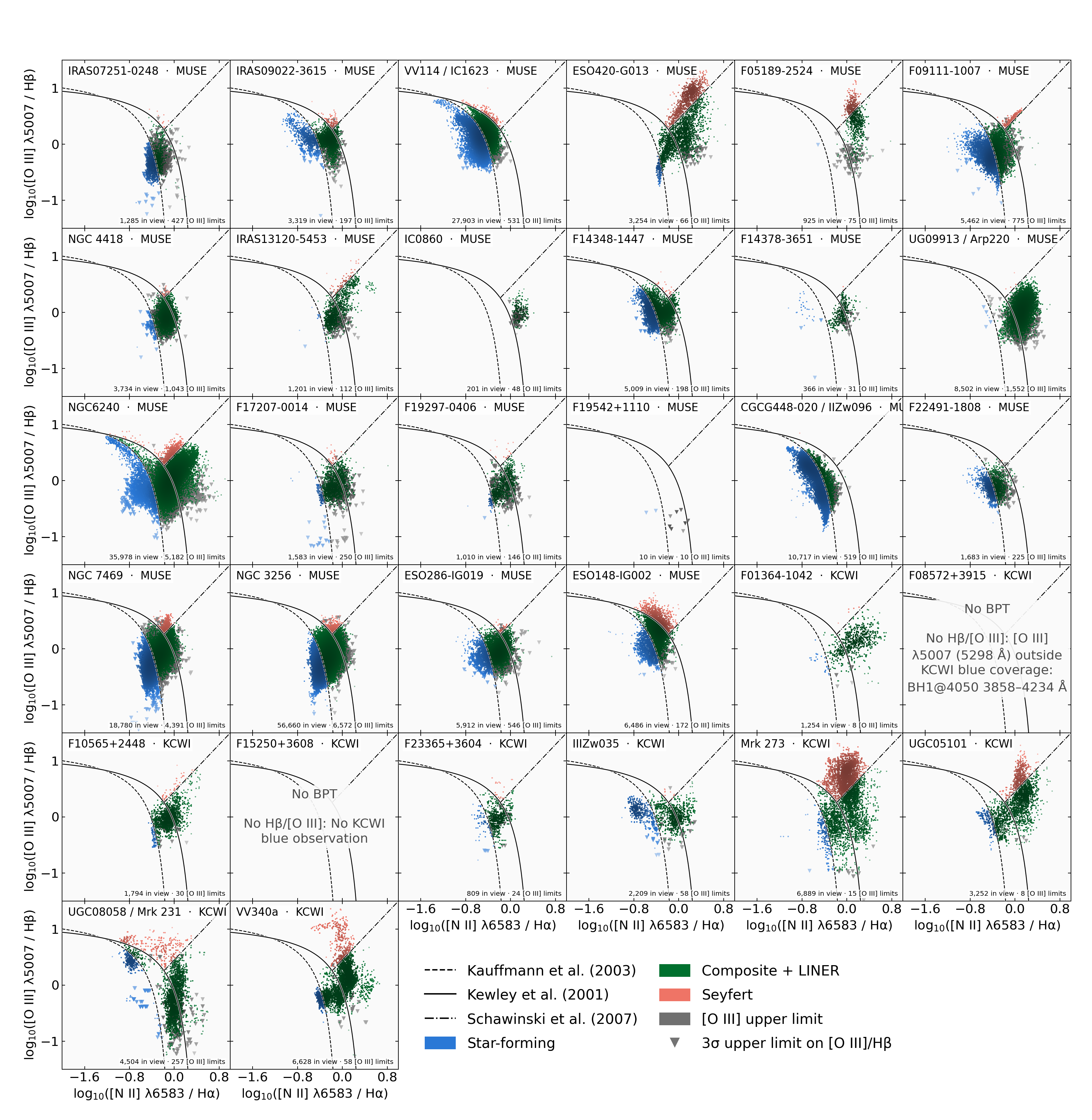}
    \caption{Resolved BPT diagrams for the 30 systems with measurements of the required lines: \oiii~$\lambda$5007/\hb\ against \nii~$\lambda$6583/H$\alpha$, one point per spaxel, for every spaxel passing the joint S/N~$\geq3$ mask in H$\alpha$, \hb\ and \nii~$\lambda$6583. Panels are labeled with the source name and instrument and share common axes; spaxels with ratios outside the plotted range are counted in the tally at the lower right of each panel, which also gives the number of spaxels in view and how many carry only an \oiii\ upper limit.  Color encodes the BPT class and shade the local density of points, estimated by a Gaussian kernel and scaled logarithmically within each panel, so that pale and dark tones of one color distinguish sparse from crowded regions of a single galaxy and cannot be compared between panels.  Spaxels undetected in \oiii~$\lambda$5007 are plotted as $3\sigma$ upper limits (triangles) rather than discarded, since requiring a detection would preferentially remove star-forming spaxels; they are classified where the limit alone places them below the \cite{Kauffmann03} line and are otherwise counted as undetermined.  Composite and LINER spaxels share a color, for the reason given in Section \ref{sec:BPT}.  The two systems without panels are IRAS F08572$+$3915 and F15250$+$3608, whose only KCWI blue setting (BH1) covers neither \hb\ nor \oiii.}
    \label{fig03:bpt}
\end{figure*}

The BPT diagrams shown in Figure \ref{fig03:bpt} contain the ratios \oiii~5007/H$\beta$ against \nii~6583/H$\alpha$ for every spaxel in the joint S/N mask defined in Section \ref{sec:fits}. Spaxels undetected in \oiii~are treated as 3$\sigma$~limits and are only classified when the limit places them below the \cite{Kauffmann03} line, and are otherwise counted as undetermined. Spaxels with undetermined BPT classification are excluded from the class fractions and carry no \oiii~weight. The median proportion of undetermined spaxels across the sample was 7\%; these limits are shown as triangles in Figure~\ref{fig03:bpt} and enter Table~\ref{tab:derived} only through the classification. 

Three boundaries are overlaid on the BPT diagrams, dividing the diagnostic plane into five regions. Below the \cite{Kauffmann03} pure star-formation boundary, points are classified as ``star-forming'' (blue). Between the \cite{Kauffmann03} line and the \cite{Kewley01} maximum starburst line, spaxels are classified as composite (green). Spaxels in the region beyond the \cite{Kewley01} line and above the \cite{schawinski07} Seyfert-LINER demarcation line are classified as Seyfert (red), while below this line they are classified as LINER (green). Composite and LINER spaxels share a single color: neither $f_{\rm{LIN}}$ nor $f_{\rm{comp}}$~correlates with $f_{\rm AGN,bol}$ (Spearman $\rho=+0.08$ and $-0.10$, $p>0.5$), whereas the Seyfert fraction does ($\rho=+0.55$, $p=0.002$), so the two are not distinguished in a map whose purpose is to locate AGN-excited gas.
Both classes are classified and tabulated separately in Table~\ref{tab:derived}.

The per-source excitation budget divides the number of star-forming, composite, LINER, and Seyfert-like spaxels by $N_{\rm{spax}}$, giving $f_{\rm{SF}}$, $f_{\rm{comp}}$, $f_{\rm{LIN}}$, and $f_{\rm{Sy}}$, respectively. These quantities are listed in Table \ref{tab:derived}. The median fractions of classified spaxels are $f_{\rm SF}=0.11$, $f_{\rm comp}=0.44$, $f_{\rm LIN}=0.26$ and $f_{\rm Sy}=0.02$ --- composite and LINER-like excitation dominate the area in most systems, and Seyfert-like gas is rare except in a handful (Mrk 273, $f_{\rm Sy}=0.50$; F05189$-$2524, $0.35$; ESO 420-G013, $0.31$).

Although the class fractions in columns 6-9 of Table~\ref{tab:derived} are useful in evaluating over how much of the galaxy does the AGN dominate the excitation, they weight every spaxel equally. Therefore, the class fractions measure the area over which each excitation mechanism dominates rather than the share of line emission the mechanism produces. Because a galaxy's projected area is dominated by its faint outskirts while its line emission is dominated by its nucleus, an area fraction is a statistic that is biased toward faint gas. Hence, an area fraction is not an energy budget, and so cannot be set against, e.g., $f_{\rm{AGN,bol}}$, which is a ratio of luminosities. Weighting each spaxel by its own line luminosity addresses these limitations. Columns 10 and 11 in Table \ref{tab:derived} correspond instead to the fraction of the  H$\alpha$ and \oiii~5007 luminosity that is Seyfert-classified, i.e., $f_{\rm{Sy}}$(H$\alpha$) = $\sum L$(H$\alpha$)$_{\rm{spax, Sy}}$ / $L_{\rm{tot}}$(H$\alpha$). This is then a fraction of a luminosity budget that can be compared with $f_{\rm{AGN,bol}}$ directly. Part of the excess of the \oiii-weighted fraction is built into the classification, since the Seyfert-classified spaxels are by construction those with the highest \oiii/H$\beta$; the H$\alpha$-weighted fraction does not share this bias. In AGN-dominated systems, $f_{\rm{Sy}}$(H$\alpha$) and $f_{\rm{Sy}}$(\oiii) tend to exceed $f_{\rm{Sy}}$ substantially: the Seyfert fraction for Mrk 273 goes from 0.5 by area to 0.88 by \oiii~luminosity.

\begin{figure*}
    \centering
    \includegraphics[width=1\linewidth]{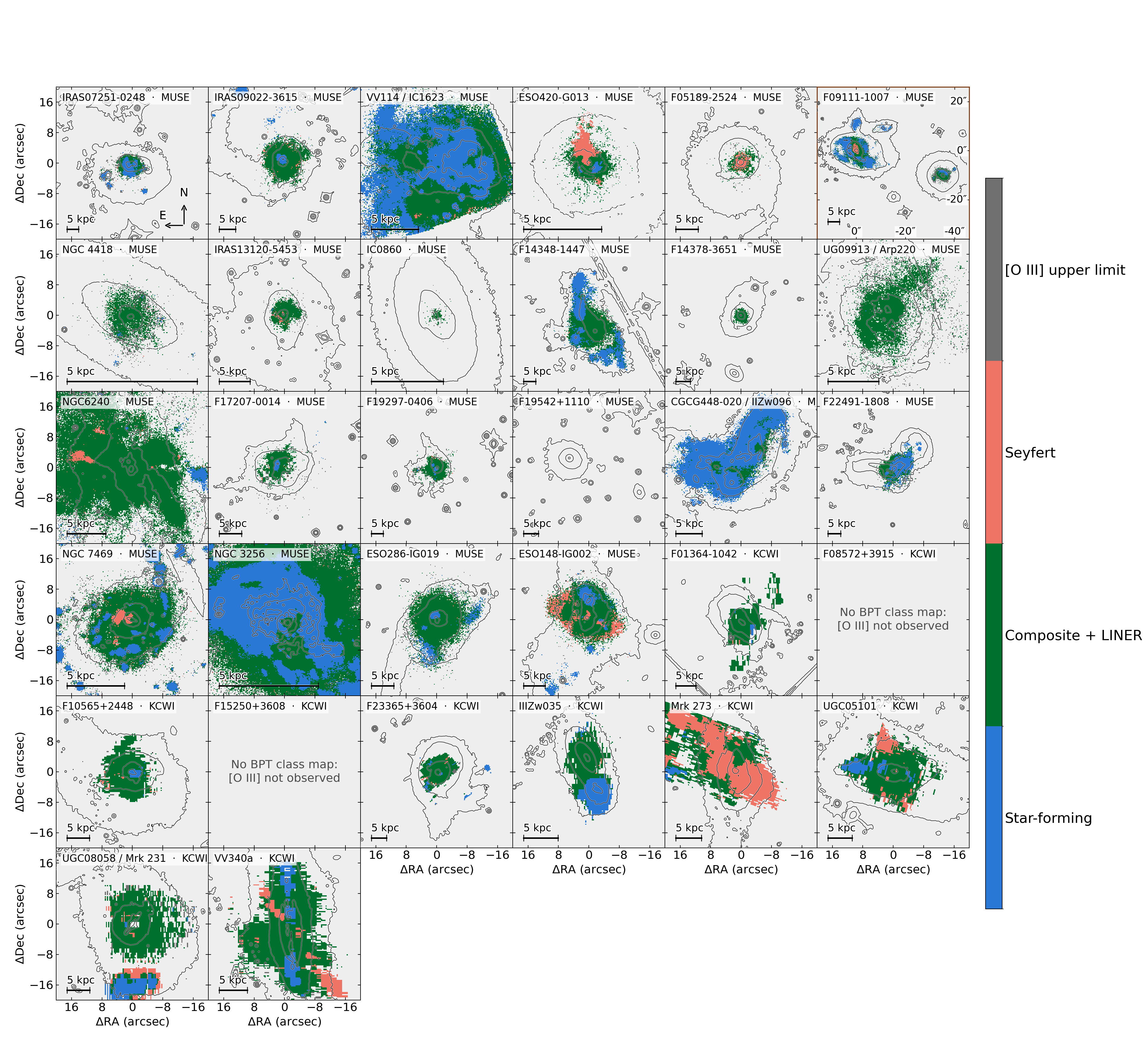}
    \caption{Maps of the BPT classification, with each spaxel of Figure \ref{fig03:bpt} colored by its class and placed at its position on the sky. Panels, fields, orientation, scale bars and \emph{HST} F814W contours are as in Figure \ref{fig01:Halpha}, and the colors are those of Figure \ref{fig03:bpt}: star-forming below the \cite{Kauffmann03} line, composite and LINER between it and the \cite{schawinski07} division, Seyfert above, and gray for spaxels whose \oiii\ upper limit leaves the class undetermined.  The morphological distinctions discussed in Section \ref{sec:BPT} are visible directly: Seyfert-like gas is nuclear or cone-like, LINER-like gas is extended and often follows tidal or outflow structure, and star forming gas traces spiral arms, discrete \hii\ complexes and starburst cores.}
    \label{fig04:bptclass}
\end{figure*}

In the BPT classification maps of Figure \ref{fig04:bptclass}, gas dominated by a given excitation mechanism tends to follow consistent morphology across the sample, and there are clear morphological distinctions between gas in different excitation regimes. Seyfert-like gas tends to be nuclear (e.g., IRAS F05189-2524) or cone-like (e.g., UGC 05101, VV340a), LINER-like gas is extended and often follows tidal or outflow structures (e.g., Arp 220, F01364-1042, VV340a), and star-forming gas closely follows spiral arms (VV 114, NGC 7469), star-forming regions (e.g., IRAS F14348-1447, II Zw 096), starburst cores (e.g., IRAS 09022-3615, IRAS F17207-0014). For the 19 ULIRGs with a mask, the median class fractions are $f_{\rm SF}=0.10$, $f_{\rm comp}=0.51$, $f_{\rm LIN}=0.26$ and $f_{\rm Sy}=0.03$, close to those of the full sample. A useful quantity is the radius of the Seyfert-dominated region $R_{\rm{Sy}} = (A_{\rm{Sy}}/\pi)^{1/2}$, where $A_{\rm{Sy}}$ is the equivalent area of the total Seyfert-classified gas, whether or not that area is contiguous. It is an area-equivalent radius rather than the radius of a circular aperture, so where the Seyfert-classified gas is offset from the nucleus (e.g., UGC 05101, NGC 6240) it measures the extent of that gas, not its distance from the nucleus. It is measured in projection and is not deconvolved from the point-spread function (Table~\ref{tab:derived}). $R_{\rm{Sy}}$ ranges from 0.08 to 7.5 kpc. This quantity is compared against AGN bolometric luminosity in Section \ref{sec:discussion} (Figure~\ref{fig05:rseyfert}).

The LINER component is not AGN-like in its behavior. $f_{\rm LIN}$ does not correlate with $f_{\rm AGN,bol}$ ($\rho=+0.08$, $p=0.66$ by area; $\rho=-0.08$, $p=0.68$ \oiii-weighted), and adding it to $f_{\rm Sy}$ destroys the correlation that $f_{\rm Sy}$ alone shows ($\rho=+0.21$, $p=0.27$). Conversely the Seyfert share of the above-Kewley gas, $f_{\rm Sy}/(f_{\rm Sy}+f_{\rm LIN})$, does correlate ($\rho=+0.49$, $p=0.008$). This suggests that in these systems the LINER-like gas is powered mainly by shocks, diffuse ionized gas and hot evolved stars rather than by the AGN, as found for the extended LINER-like emission of GOALS mergers \citep{Rich11,Rich15}, and it is the justification for keeping the \cite{schawinski07} split rather than merging the two classes.

\section{Discussion}
\label{sec:discussion}

\begin{figure*}
    \centering
    \includegraphics[width=1\linewidth]{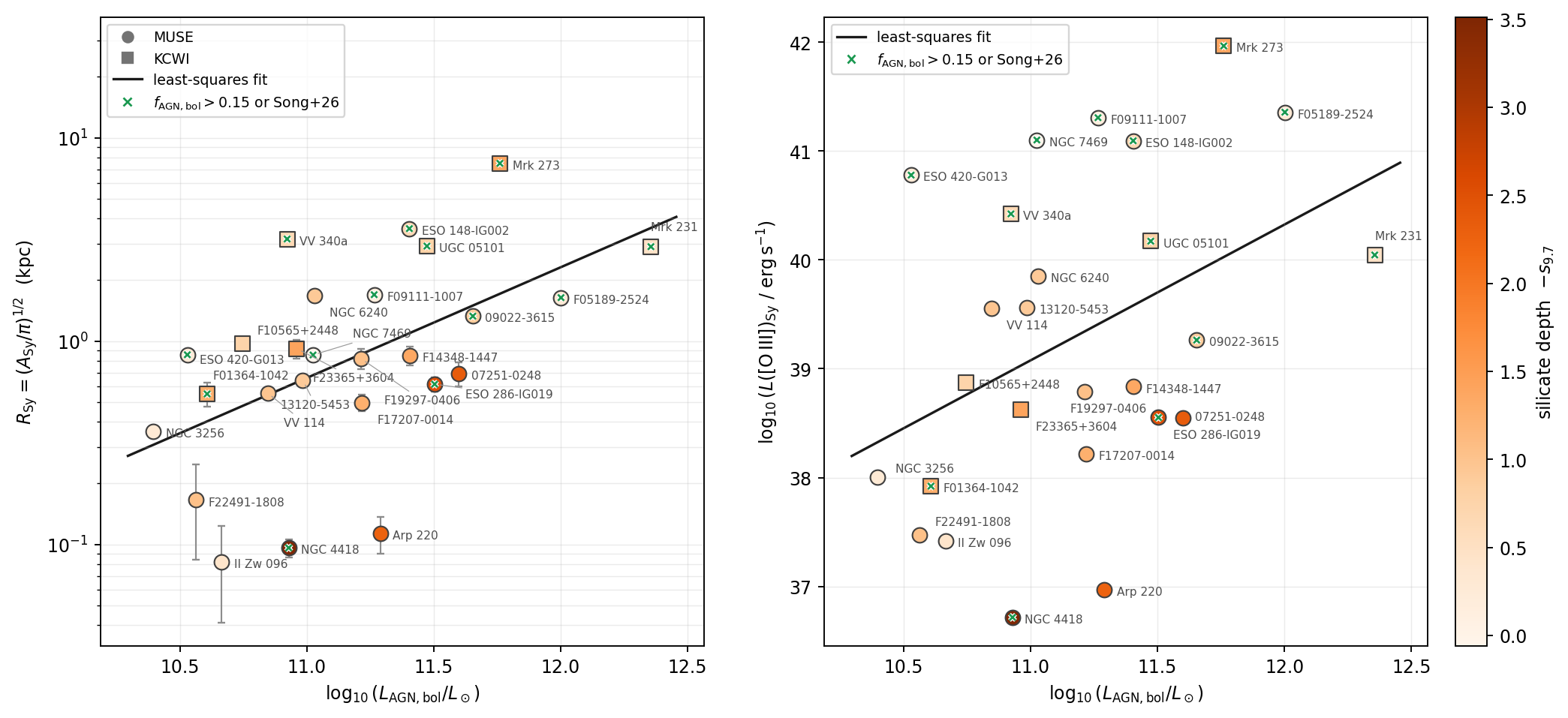}
    \caption{The Seyfert-classified gas scales with the luminosity of the AGN that could excite it.  \emph{(a)} $R_{\rm Sy}=(A_{\rm Sy}/\pi)^{1/2}$, the equivalent radius of the Seyfert-classified area, against $L_{\rm AGN,bol}=1.15\sum_i f_{{\rm AGN},i}L_{{\rm IR},i}$ over the nuclei inside the IFU field (Table \ref{tab:sample}).  \emph{(b)} $L(\oiii)$ summed over the same spaxels, also against $L_{\rm AGN,bol}$; no area enters this panel. Circles are MUSE and squares KCWI, and both panels share the color scale at right, which encodes the $9.7$\,$\mu$m silicate depth of \cite{Stierwalt13}.  A green cross marks the systems with $f_{\rm AGN}>0.15$, and IRAS F01364$-$1042, whose obscured AGN was identified by \cite{song26}.  Vertical error bars are the spaxel bootstrap described in Section \ref{sec:discussion}. The solid lines are ordinary least-squares fits, with slopes $0.54\pm0.18$ in (a) and $1.25\pm0.57$ in (b).  Only the 27 systems with at least 500 classified spaxels are considered; 26 of them appear, since III Zw 035 has $f_{\rm AGN}=0.00$ \citep{Diaz-Santos17} and so no defined $L_{\rm AGN,bol}$.}
    \label{fig05:rseyfert}
\end{figure*}

\begin{figure}
    \centering
    \includegraphics[width=1\linewidth]{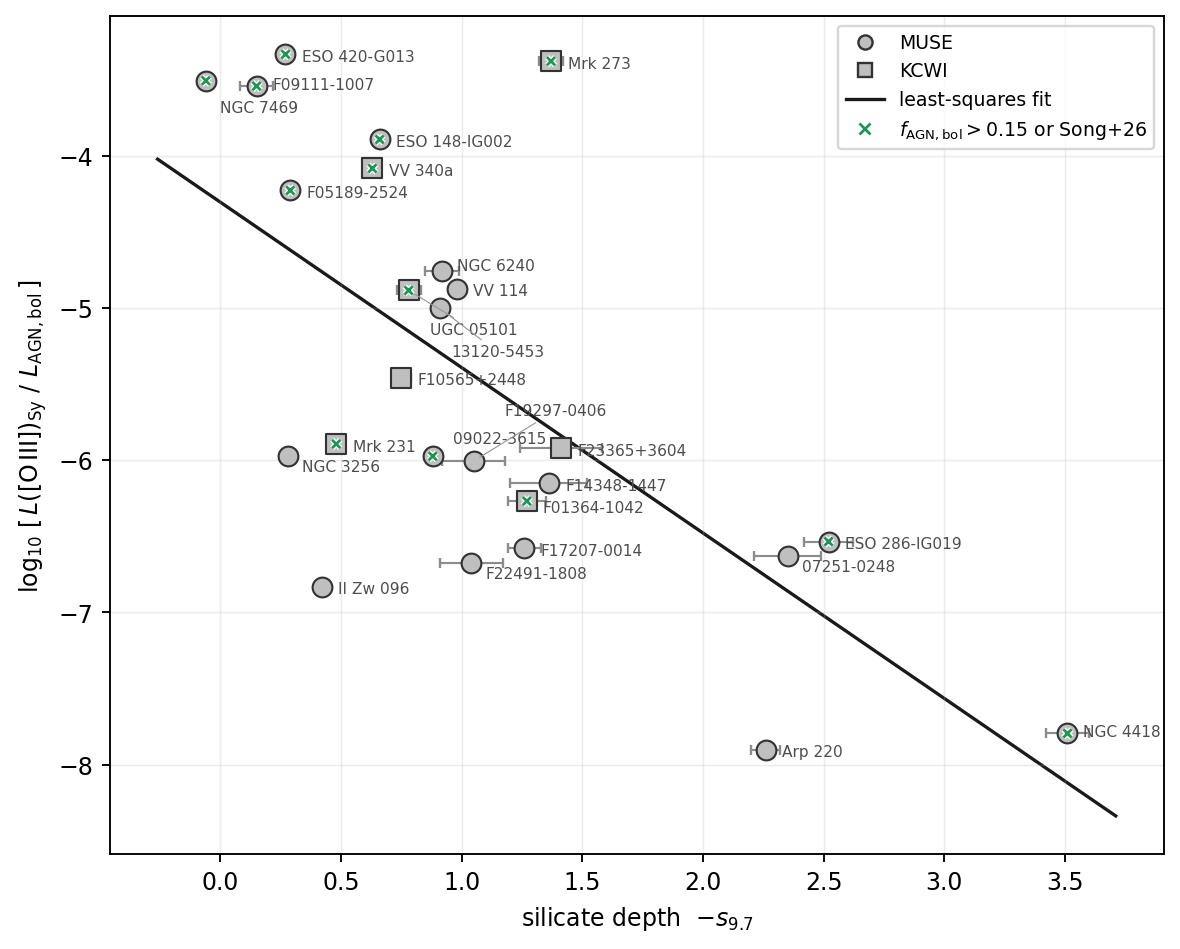}
    \caption{The \oiii\ output of the Seyfert-classified gas per unit AGN
    bolometric luminosity, against the depth of the $9.7$\,$\mu$m silicate absorption feature. The solid line is an ordinary least-squares fit with slope $-1.09\pm0.25$ dex per unit depth ($p=0.0002$, $r^{2}=0.45$); Spearman $\rho=-0.60$ ($p=0.0011$, $N=26$).  Symbols and green crosses as in Figure \ref{fig05:rseyfert}. The most obscured nuclei (NGC 4418, Arp 220 and IRAS 07251$-$0248) return two to three orders of magnitude less \oiii\ per unit AGN luminosity than the least obscured. Figure~\ref{fig:app-merger} shows the same diagram colored by merger stage.}
    \label{fig06:deficit}
\end{figure}

The classification maps of Section \ref{sec:BPT} show where the excitation is AGN-like, but not whether that gas is ionized by the AGN known from the mid-infrared to be present in some of the systems. MAUNA offers an independent handle on the second question: every system has a mid-IR estimate of the share of the total bolometric luminosity attributable to the AGN \citep{Diaz-Santos17}, derived from diagnostics far less sensitive to extinction than optical measures, and therefore able to see an AGN that the optical classification cannot. In this section, the question of whether the Seyfert-classified gas scales with that independently measured AGN power is addressed, and what governs the scatter in cases where there is a correlation. The gas is characterized in two ways, by its size and by its luminosity. The equivalent radius $R_{\rm{Sy}}$ of the Seyfert-classified area (which was defined in Section \ref{sec:BPT}) and the \oiii~luminosity $L(\oiii)_{\rm{Sy}}$~summed over the same spaxels. The two characterizations are complementary in that they are not sensitive to the same systematics, although they are measured over the same spaxels (the geometric quantity is sensitive to projection and the S/N mask definition; the energetic quantity is sensitive to extinction and absolute flux calibration). Both quantities are found to scale with $L_{\rm{AGN,bol}}$ (Figure \ref{fig05:rseyfert}). The scatter about the scaling is large but correlates with the depth of the 9.7 $\mu$m silicate absorption (Figure \ref{fig06:deficit}).

The vertical error bars in Figure \ref{fig05:rseyfert} are obtained by resampling with replacement the spaxels in each galaxy's mask
(500 draws) and recomputing the quantity for each draw; the quoted uncertainty is the standard deviation over the draws.  For $R_{\rm Sy}$, which scales as $N_{\rm Sy}^{1/2}$, the fractional uncertainty is therefore half that on the Seyfert spaxel count. These uncertainties describe the counting statistics of a finite mask and nothing else: they do not include the systematic effect of
spaxels near a classification boundary, nor the absolute flux calibration.

The solid lines in Figure \ref{fig05:rseyfert} are ordinary least-squares fits. In panel (a), $\log R_{\rm Sy} = (0.54\pm0.18)\log L_{\rm AGN,bol} - (6.17\pm1.98)$ ($r^2=0.28$), consistent with the slope of $1/2$ expected for gas photoionized at a constant ionization parameter $U\propto L/(R^2 n)$ at fixed density. An ionization-bounded Str\"omgren sphere at fixed density would instead give $R\propto L^{1/3}$. \cite{Bennert02} found $R\propto L^{0.52}$ for the narrow-line regions of Seyferts and quasars, while samples of luminous obscured quasars give flatter relations \citep{Greene11,Hainline13}. In panel (b) the slope is $1.25\pm0.57$ ($r^2=0.17$), consistent with $L(\oiii)_{\rm Sy}\propto L_{\rm AGN,bol}$.

The left panel of Figure \ref{fig05:rseyfert} shows that the size of the Seyfert region tracks AGN luminosity (Spearman $\rho=+0.50$; $p=0.010$; $N=26$). Whether a third variable produces the correlation is tested with rank partial correlation: the ranks of both quantities are regressed on the ranks of the control variable, and the Spearman correlation of the residuals is quoted. In a flux-limited sample both axes could grow with distance, but with the corrected distances $R_{\rm Sy}$ does not correlate with luminosity distance ($\rho=+0.15$, $p=0.45$), and with distance controlled the correlation is unchanged ($\rho=+0.48$, $p=0.016$). Controlling for $L_{\rm IR}$, which enters $L_{\rm AGN,bol}$, strengthens it ($\rho=+0.55$, $p=0.004$). The correlation is more sensitive to spatial resolution: with the physical size of the point-spread function controlled (for MUSE the delivered image quality, for KCWI the seeing combined with the slice width) it falls to $\rho=+0.39$ ($p=0.055$), and restricting the sample to systems whose Seyfert region is resolved and not truncated by the field edge weakens it further (Appendix~\ref{sec:app-robust}). Panel (a) is therefore a size--luminosity relation that resolution partly amplifies. In the right panel, $L(\oiii)_{\rm Sy}$ correlates with $L_{\rm AGN,bol}$ ($\rho=+0.39$; $p=0.051$), and controlling for distance ($\rho=+0.44$, $p=0.026$) or for $L_{\rm IR}$ ($\rho=+0.54$, $p=0.005$) strengthens it; but on its own this relation is marginal.

The two correlations in Figure \ref{fig05:rseyfert} use different properties of the same Seyfert-classified spaxels, their number in one case and their \oiii\ luminosity in the other, and so are complementary (but not independent) tests of whether the optical line-emitting gas traces the energy budget of the AGN.

Figure \ref{fig06:deficit} shows that the scatter in the $L(\oiii)_{\rm Sy}$--$L_{\rm AGN,bol}$ relation is governed by nuclear obscuration. The ratio $L(\oiii)_{\rm Sy}/L_{\rm AGN,bol}$ spans 4.6 dex and falls with silicate depth ($\rho = -0.60$; $p = 0.0011$; $N = 26$); the solid line is a least-squares fit with slope $-1.09\pm0.25$ dex per unit depth ($r^2 = 0.45$). Equivalently, $L(\oiii)_{\rm Sy}$ against silicate depth at fixed $L_{\rm AGN,bol}$ gives $\rho = -0.69$ ($p = 0.0001$), and controlling for distance leaves $\rho=-0.58$ ($p=0.003$). The same test on size, $\log R_{\rm Sy}-\tfrac12\log L_{\rm AGN,bol}$ against depth, gives a weaker anticorrelation ($\rho = -0.39$; $p = 0.047$): the obscured systems are somewhat smaller in projection, but much fainter in \oiii.

The deficit could arise from dust absorbing the ionizing photons before they reach the gas, from dust hiding the ionized gas along our line of sight, or from a genuinely smaller ionized region. For the gas that is detected, the Balmer decrement distinguishes the first two (Appendix~\ref{sec:app-robust}): the extinction of the Seyfert-classified gas (median $A_V\approx2$ mag) does not correlate with silicate depth ($\rho=-0.10$) and is an order of magnitude below the screen implied by the silicate feature, and correcting \oiii\ for it spaxel by spaxel leaves the trend intact ($\rho=-0.57$) and steepens it (slope $-1.43\pm0.37$; Figure \ref{fig:app-balmer}). In many systems the Seyfert-classified spaxels also lie outside the nuclear regions (Figure~\ref{fig04:bptclass}); if the intrinsic \oiii\ is brightest at the AGN, much of the nuclear \oiii\ is already missing from our measurements. The missing \oiii\ is therefore not emitted by visible gas behind a foreground screen; it comes from gas that is hidden entirely or that the obscured nucleus does not ionize, and a correction of the observed \oiii\ based on the silicate depth would not be appropriate for the gas the optical data detect. The decisive test is the mid-infrared fine-structure lines measured with \emph{JWST}/MIRI (e.g., [Ne\,\textsc{v}]\,14.3, 24.3\,$\mu$m and [O\,\textsc{iv}]\,25.9\,$\mu$m), which are far less sensitive to dust, and with the MRS cubes this comparison can be made spatially resolved. The \emph{Spitzer} [O\,\textsc{iv}] fluxes of \cite{Inami13} cannot provide the test: every system deeper than $-s_{9.7}\approx1.4$ has only an upper limit, and the \oiii/[O\,\textsc{iv}] ratio shows no significant trend with depth (Appendix~\ref{sec:app-robust}).

The trend is not produced by the construction of $f_{\rm AGN}$. Because $f_{\rm AGN}$ and the silicate depth are measured from the same IRS spectra, and PAH equivalent width co-varies with silicate depth \citep{Spoon07}, an $f_{\rm AGN}$ biased high in obscured nuclei could create the trend. In our sample, however, $f_{\rm AGN}$ does not correlate with silicate depth ($\rho=-0.14$, $p=0.48$): its individual diagnostics do, in opposite directions (the 30-to-15\,$\mu$m continuum ratio, $\rho=-0.53$; the PAH equivalent width, $\rho=+0.34$), and the average cancels. The $L(\oiii)_{\rm Sy}/L_{\rm AGN,bol}$\ vs. silicate depth trend is significant in 11 of 13 sample variants (Appendix~\ref{sec:app-robust}), including without NGC 4418 ($\rho=-0.56$, $p=0.004$), without both NGC 4418 and Arp 220, whose AGN content is debated ($\rho=-0.51$, $p=0.011$), and for the 20 systems whose Seyfert region is resolved and not truncated by the field ($\rho=-0.66$, $p=0.002$); it is significant in 87\% of Monte Carlo realizations that perturb every quantity within its uncertainty and resample the systems. It also holds with \oiii\ summed over all classified gas rather than the Seyfert-classified gas alone ($\rho=-0.48$), so it does not depend on the classification. Later merger stages lie preferentially below the fit (residual against merger class: $\rho=-0.50$, $p=0.010$; Figure~\ref{fig:app-merger}), although 18 of the 26 systems are late-stage mergers.

The largest positive residual is Mrk 273, whose Seyfert-classified gas extends to 7.5\,kpc despite a moderately deep silicate feature, consistent with ionizing radiation escaping along an optically thin direction such as its outflow. The largest negative residuals are II Zw 096, IRAS F22491$-$1808 and Arp 220, which have one, one and six Seyfert-classified spaxels respectively, so that their $L(\oiii)_{\rm Sy}$ rests on very little gas.

Because MAUNA is weighted toward luminous, late-stage mergers (Section~\ref{sec:sample}), these trends describe that population; whether they hold for earlier-stage, lower-luminosity LIRGs needs a sample matched to GOALS. These results extend earlier evidence that optical spectra miss much of the AGN power in infrared-luminous galaxies \citep{Goulding09,Yuan10}. Figure \ref{fig06:deficit} adds a quantitative, spatially resolved form of the same effect: Seyfert-classified gas is still found in the obscured systems, but the \oiii\ luminosity it carries per unit AGN power declines by an order of magnitude per unit silicate depth.

\section{Summary and Conclusions}
\label{sec:summary_conclusions}

We have presented resolved optical integral-field spectroscopy of the 32
(U)LIRG systems of the MAUNA sample, drawn from GOALS and selected for having
both optical IFU coverage and approved or completed \emph{JWST}/MIRI
observations.  The data comprise archival VLT/MUSE cubes for 22 systems and new
Keck/KCWI observations obtained over ten nights between 2020 and 2024, 12 of the
MUSE targets having KCWI coverage as well.  Every cube was fitted spaxel by
spaxel with \textsc{HyperCube} using a common emission-line model over a stellar
continuum, and all maps, diagrams and derived quantities in this paper rest on a
single, uniformly applied signal-to-noise mask.  Our main results are:

\begin{enumerate}

\item Composite and LINER-like excitation together account for most of the classified area (Section~\ref{sec:BPT}; Table~\ref{tab:derived}). The median excitation budget over the classified spaxels is $f_{\rm SF}=0.10$, $f_{\rm comp}=0.44$, $f_{\rm LIN}=0.26$ and $f_{\rm Sy}=0.02$.  Seyfert-like gas is rare in area terms except in a handful of systems (Mrk 273, $f_{\rm Sy}=0.50$; IRAS F05189$-$2524, $0.35$; ESO 420-G013, $0.31$), although it carries far more of the line luminosity than of the area wherever it is present, e.g., in Mrk 273 the share of Seyfert-like gas rises from $0.50$ by area to $0.88$ by \oiii\ luminosity, partly because Seyfert-classified spaxels have by construction the highest \oiii/H$\beta$.

\item The LINER-like gas does not appear to be AGN-powered (Section~\ref{sec:BPT}). Its areal fraction does not correlate with the mid-IR AGN fraction ($\rho=+0.08$, $p=0.66$), and combining it with the Seyfert fraction destroys the correlation that $f_{\rm{Sy}}$ shows on its own. The Seyfert share of the gas above the \cite{Kewley01} line does correlate with $f_{\rm AGN,bol}$ ($\rho=+0.49$, $p=0.008$). In these merger-dominated systems the LINER-like emission is more plausibly attributed to shocks, diffuse ionized gas and hot evolved stars than to the AGN.

\item The Seyfert-classified gas scales with AGN power, in both size and luminosity (Section~\ref{sec:discussion}; Figure~\ref{fig05:rseyfert}). The equivalent radius of the Seyfert-classified region correlates with the bolometric AGN luminosity ($\rho=+0.50$, $p=0.010$, $N=26$), with $R_{\rm Sy}\propto L_{\rm AGN,bol}^{0.54\pm0.18}$ as expected at constant ionization parameter, and the \oiii\ luminosity of the same spaxels follows it marginally ($\rho=+0.39$, $p=0.051$).  Both survive control for luminosity distance ($\rho=+0.48$ and $+0.44$) and for $L_{\rm IR}$ ($+0.55$ and $+0.54$); the size relation is partly amplified by resolution, falling to $\rho=+0.39$ when the physical size of the point-spread function is controlled. The two tests are complementary: one uses only the area and the other only the luminosity of the same spaxels. $R_{\rm Sy}$ ranges from $0.08$ to $7.5$\,kpc, with 28 of the 30 systems with a mask containing at least one Seyfert-classified spaxel.

\item Nuclear obscuration governs the scatter (Section~\ref{sec:discussion}; Figure~\ref{fig06:deficit}; Appendix~\ref{sec:app-robust}). At fixed $L_{\rm AGN,bol}$ the \oiii\ luminosity of the Seyfert-classified gas spans $4.6$ dex, and that scatter is not random: the ratio $L(\oiii)_{\rm Sy}/L_{\rm AGN,bol}$ falls with the depth of the $9.7$\,$\mu$m silicate feature at $-1.09\pm0.25$ dex per unit depth ($\rho=-0.60$, $p=0.0011$).  The most obscured nuclei in the sample (NGC 4418, Arp 220 and IRAS 07251$-$0248) return two to three orders of magnitude less \oiii\ per unit AGN luminosity than the least obscured systems. The trend survives controls for distance, $L_{\rm IR}$, resolution and instrument, the removal of NGC 4418 and Arp 220, and Monte Carlo perturbation of every quantity, and it is not caused by foreground extinction of the visible gas, whose Balmer decrement is unrelated to silicate depth. The equivalent test on the size of the region is weaker ($\rho=-0.39$, $p=0.047$), so the obscured systems are fainter in \oiii\ much more than they are smaller.

\end{enumerate}

These results establish a foundation for the work MAUNA was intended for. The same spaxels that Figure \ref{fig04:bptclass} classifies carry kinematic information that a forthcoming paper will use to separate rotating gas from outflowing gas, and the \oii~3726,29 doublet measured by the KCWI blue arm extends the density and ionization diagnostics to the low-ionization structures that \oiii\ alone cannot trace.  Most directly, the obscuration trend in Figure \ref{fig06:deficit} is a prediction that the \emph{JWST}/MIRI observations of these same nuclei can test: if the \oiii\ deficit in the most obscured systems is an extinction effect, the mid-infrared lines, which are far less sensitive to dust, should show no corresponding deficit at fixed $L_{\rm AGN,bol}$, and the MRS cubes will allow the optical and mid-infrared diagnostics to be compared spaxel by spaxel. Far-infrared spectroscopy with the PRIMA probe \citep{glenn25} will extend these dust-insensitive diagnostics to longer wavelengths.

\section{Data Release}
\label{sec:data_release}

All of the original KCWI calibration and science exposures used here are from Keck programs 2020B\_U055, 2021B\_U134, 2021B\_K318, 2022A\_U213, 2022A\_K302, 2023A\_E327, 2023B\_K454, 2023B\_U081, 2024A\_U305 and 2024A\_K453 and are publicly available on the Keck Observatory Archive (KOA\footnote{https://koa.ipac.caltech.edu/cgi-bin/KOA/nph-KOAlogin}). The \textsc{HyperCube} code used for the fits is archived on Zenodo \citep{Kader26_HyperCube}. The archival MUSE cubes are available from the ESO Science Archive Facility\footnote{https://archive.eso.org/scienceportal/home} (Table~\ref{tab:muse-provenance}). All the \emph{HST} and \emph{JWST} data used in this paper can be found in MAST\footnote{https://mast.stsci.edu/portal/Mashup/Clients/Mast/Portal.html}: \dataset[10.17909/j5bq-dd62]{https://doi.org/10.17909/j5bq-dd62}.

\begin{acknowledgments}

JK and VU gratefully acknowledge partial funding support from National Science Foundation (NSF) Astronomy and Astrophysics Research Grant \#2536603, NASA ADSPS grant \#80NSSC25K0169, as well as STScI grant \#JWST-GO-08391.001-A, which was provided by NASA through a grant from the Space Telescope Science Institute, which is operated by the Association of Universities for Research in Astronomy, Inc., under NASA contract.
This work was performed in part at the Aspen Center for Physics, which is supported by National Science Foundation grant PHY-2210452.
TG acknowledges support from the Australian Research Council through Discovery Project DP210101945, funded by the Australian Government.
JNS acknowledges support provided by the NSF through the Grote Reber Fellowship Program administered by Associated Universities, Inc./National Radio Astronomy Observatory.

Some of the data presented herein were obtained at Keck Observatory, which is a private 501(c)3 non-profit organization operated as a scientific partnership among the California Institute of Technology, the University of California, and the National Aeronautics and Space Administration. The Observatory was made possible by the generous financial support of the W. M. Keck Foundation. The authors wish to recognize and acknowledge the very significant cultural role and reverence that the summit of Maunakea has always had within the Native Hawaiian community. We are most fortunate to have the opportunity to conduct observations from this mountain.

This research has made use of the Keck Observatory Archive (KOA), which is operated by the W. M. Keck Observatory and the NASA Exoplanet Science Institute (NExScI), under contract with the National Aeronautics and Space Administration.

Based on observations collected at the European Southern Observatory under ESO programs 60.A-9100(H), 60.A-9339(A), 095.B-0049(A), 095.B-0482(A), 097.B-0165, 097.B-0427(A), 099.B-0426(A), 0101.B-0368(B), 0102.B-0617(A), 0103.B-0391(A), 0103.B-0927(A), 0104.B-0151(A), 0104.B-0668(A,B,D) and 111.253U.001.

This work is based in part on observations made with the NASA/ESA/CSA \emph{James Webb Space Telescope} and the NASA/ESA \emph{Hubble Space Telescope}, obtained from the Mikulski Archive for Space Telescopes (MAST) at the Space Telescope Science Institute, which is operated by the Association of Universities for Research in Astronomy, Inc., under NASA contracts NAS 5-03127 for \emph{JWST} and NAS 5-26555 for \emph{HST}. The \emph{JWST} observations are associated with programs 1204, 1265, 1267, 1268, 1328, 1717, 1875, 1991, 3368 and 3869, and the \emph{HST} observations with programs 10592, 14715, 14728 and 16914. This work is based in part on archival data obtained with the \emph{Spitzer Space Telescope}, which was operated by the Jet Propulsion Laboratory, California Institute of Technology, under a contract with NASA.

\end{acknowledgments}

%

\vspace{5mm}
\facilities{KeckII:KCWI, VLT:MUSE, HST:ACS, HST:WFC3, JWST:MIRI, Spitzer:IRS}

\software{\textsc{HyperCube} \citep[v0.5.1;][]{Kader26_HyperCube},
KCWI\_DRP \citep[v1.3.1;][]{morrisey18},
CWITools \citep{osullivan20},
pPXF \citep{Cappellari17}, E-MILES \citep{Vazdekis16}, LMFIT, astropy \citep{astropy22}. 
\newline
\emph{Disclosure of Delegation to Generative AI:} the authors declare the use of generative AI (GAI) in the research and writing process. According to the GAIDeT taxonomy (2026), the following tasks were delegated to GAI tools under full human supervision (1) Code generation; (2) Process automation; (3) Creation of algorithms for data analysis; (4) Data cleaning; (5) Visualization; (6) Text generation; (7) Proofreading and editing; and (8) Monitoring compliance with ethical standards. The GAI tool used was: Claude Opus 5.5. Responsibility for the final manuscript lies entirely with the authors. GAI tools are not listed as authors and do not bear responsibility for the final outcomes. Declaration submitted by: Justin Kader}

\appendix

\section{Provenance of the archival MUSE data}
\label{sec:app-muse}

All MUSE observations used here are archival.  Table \ref{tab:muse-provenance} lists, for each source, the ESO program that produced the cube adopted in this work, its principal investigator, and whether the product is a MUSE-DEEP combined cube or a standard pipeline cube.  The distinction matters for the depth of the mask: the MUSE-DEEP products combine multiple observing blocks and reach fainter surface brightness than the standard products, so the number of spaxels passing the signal-to-noise cut in Table \ref{tab:derived} is a property of the effective instrument sensitivity in addition to an intrinsic property of the galaxy.

\begin{deluxetable*}{llccl}
\tabletypesize{\scriptsize}
\tablewidth{0pt}
\tablecaption{Provenance of the archival MUSE data used in this work.
``DEEP'' denotes an ESO MUSE-DEEP combined product and ``Std'' a standard
MUSE pipeline cube; these are the same labels used in the Prod.\ column of
Table~\ref{tab:ifu-dataset}.  PID is the ESO observing program identifier,
and PI the principal investigator of that program.
\label{tab:muse-provenance}}
\tablehead{
\colhead{IRAS ID} & \colhead{Alt.\ Name} & \colhead{Product} &
\colhead{PI} & \colhead{PID}
}
\startdata
07251-0248 & \nodata & DEEP & S. Arribas & 0104.B-0151(A) \\
09022-3615 & \nodata & DEEP & S. Arribas & 0104.B-0151(A) \\
F01053-1746 & VV 114 & DEEP & G. C. Privon & 097.B-0427(A) \\
F04118-3207 & ESO 420-G013 & DEEP & R. Amor\'{i}n & 111.253U.001 \\
F05189-2524 & \nodata & DEEP & S. Arribas & 0104.B-0151(A) \\
F09111-1007 & \nodata & Std & G. C. Privon & 099.B-0426(A) \\
F12243-0036 & NGC 4418 & DEEP & F. Stanley & 0104.B-0668(A) \\
13120-5453 & \nodata & Std & A. Fl\"utsch & 0102.B-0617(A) \\
F13126+2453 & IC 0860 & Std & F. Stanley & 0104.B-0668(B) \\
F14348-1447 & \nodata & DEEP & A. Fl\"utsch & 0102.B-0617(A) \\
F14378-3651 & \nodata & Std & A. Fl\"utsch & 0102.B-0617(A) \\
F15327+2340 & Arp 220 & DEEP & P. van der Werf & 0103.B-0927(A) \\
F16504+0228 & NGC 6240 & DEEP & E. Treister & 095.B-0482(A) \\
F17207-0014 & \nodata & DEEP & F. Stanley & 0104.B-0668(D) \\
F19297-0406 & \nodata & DEEP & S. Arribas & 0103.B-0391(A) \\
F19542+1110 & \nodata & DEEP & S. Arribas & 0103.B-0391(A) \\
F20550+1655 & II Zw 096 & DEEP & G. C. Privon & 097.B-0427(A) \\
F22491-1808 & \nodata & Std & R. Maiolino & 0101.B-0368(B) \\
F23007+0836 & NGC 7469 & Std & MUSE SV team & 60.A-9339(A) \\
F10257-4339 & NGC 3256 & Std & C.M. Carollo & 097.B-0165 \\
F20551-4250 & ESO 286-IG019 & Std & K. Soto & 095.B-0049(A) \\
F23128-5919 & ESO 148-IG002 & Std & MUSE team & 60.A-9100(H) \\
\enddata
\tablecomments{All MUSE data are archival and were retrieved from the ESO
Science Archive Facility.  program 095.B-0049(A) is MUSE guaranteed-time
observation (GTO) time; 60.A-9339(A) is MUSE Science Verification and
60.A-9100(H) MUSE commissioning time, neither of which has an individual PI,
so the responsible team is listed instead.  Where more than one MUSE program
covers a source, the entry gives the program that produced the specific cube
adopted here.}
\end{deluxetable*}

\section{Robustness of the correlations}
\label{sec:app-robust}

This appendix contains the tests behind the statements in Section~\ref{sec:discussion} that the correlations of Figures~\ref{fig05:rseyfert} and \ref{fig06:deficit} are not produced by the sample, the distance scale, the resolution, the construction of the mid-infrared quantities, or foreground extinction. Unless stated otherwise, correlations are Spearman rank coefficients over the 26 systems of Figure~\ref{fig06:deficit}, and controlled correlations are rank partial correlations.

\begin{figure*}
    \centering
    \includegraphics[width=1\linewidth]{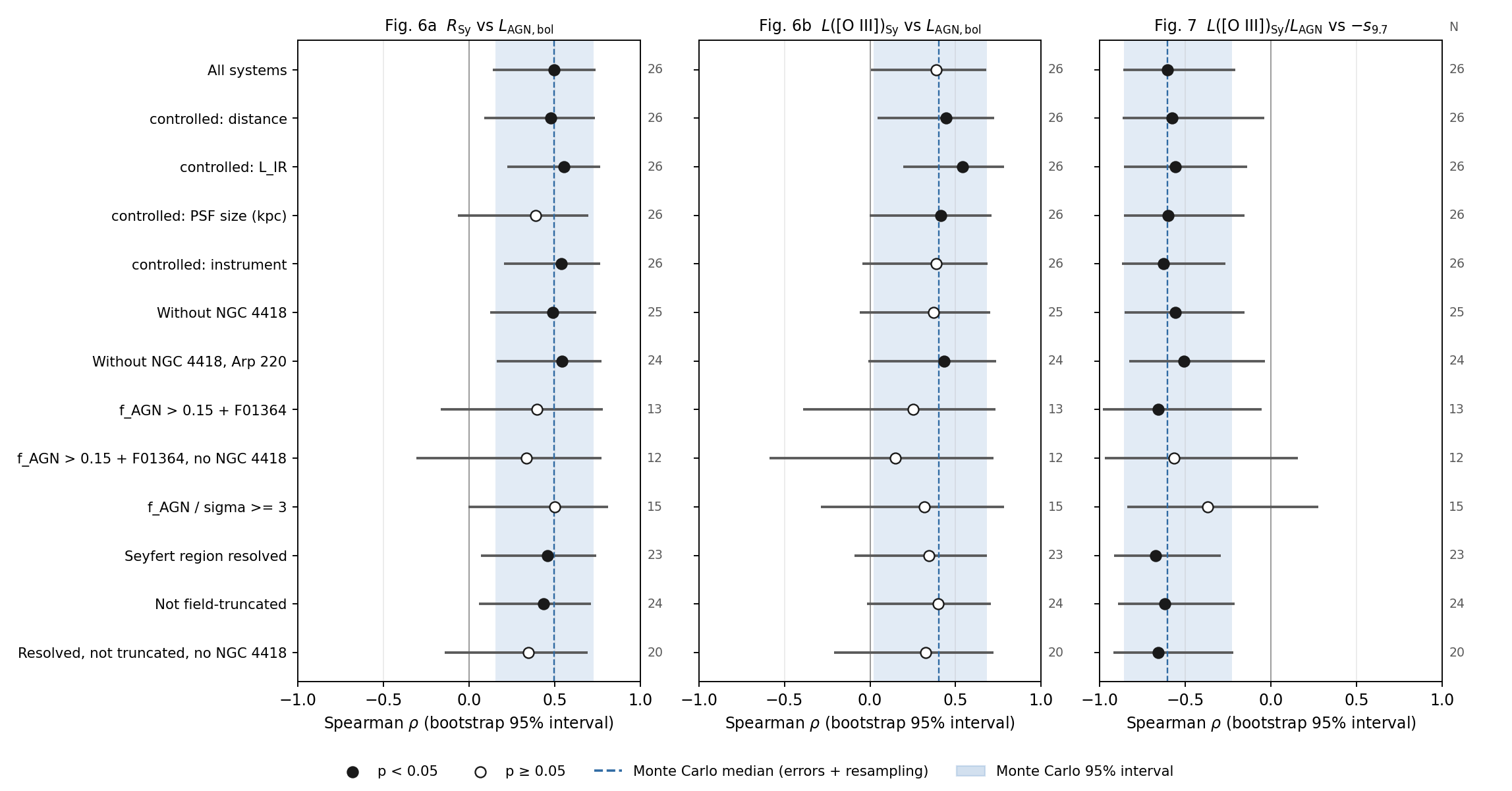}
    \caption{Robustness of the three correlations of Figures~\ref{fig05:rseyfert} and \ref{fig06:deficit}. Each row is a sample variant or control variable (Appendix~\ref{sec:app-robust}); points give the Spearman (or rank partial) correlation with its bootstrap 95\% interval, filled where $p<0.05$, and the number of systems at right. The dashed line and shaded band are the median and 95\% interval of the Monte Carlo realizations that perturb every quantity within its uncertainty and resample the systems.}
    \label{fig:app-forest}
\end{figure*}

\emph{Sample variants and controls.} Figure~\ref{fig:app-forest} shows each correlation for 13 variants: five control variables (distance, $L_{\rm IR}$, the physical size of the point-spread function, and instrument), the removal of the systems whose AGN content is debated (NGC 4418, and NGC 4418 with Arp 220), restriction to the systems with $f_{\rm AGN}>0.15$ (with IRAS F01364$-$1042 added), to those whose $f_{\rm AGN}$ is measured at $\geq3\sigma$, to those whose Seyfert region is resolved, to those not truncated by the edge of the field (VV 114 and Mrk 231 each have $\approx19\%$ of their Seyfert spaxels at the field edge), and the combination of the last two without NGC 4418. The Figure~\ref{fig06:deficit} trend is significant in 11 of the 13 variants, with $-0.67\leq\rho\leq-0.51$ in all but one. The exception is the $f_{\rm AGN}/\sigma\geq3$ cut ($\rho=-0.37$, $p=0.17$, $N=15$), which removes the deep-silicate end of the sample, where AGN fractions are least certain. The size relation of Figure~\ref{fig05:rseyfert}a is significant in 8 of 13 variants and loses significance when the size of the point-spread function is controlled ($\rho=+0.39$, $p=0.055$) or under the strictest combined cut ($\rho=+0.34$, $p=0.14$, $N=20$). The luminosity relation of Figure~\ref{fig05:rseyfert}b is significant in 4 of 13.

\emph{Monte Carlo.} Perturbing $f_{\rm AGN}$, $-s_{9.7}$ and the optical quantities within their uncertainties (5000 realizations), the median correlations are $+0.50$ (95\% interval $+0.36$ to $+0.65$) for Figure~\ref{fig05:rseyfert}a, $+0.40$ ($+0.26$ to $+0.56$) for Figure~\ref{fig05:rseyfert}b and $-0.60$ ($-0.69$ to $-0.51$) for Figure~\ref{fig06:deficit}, significant at $p<0.05$ in 94\%, 57\% and 100\% of realizations. When the systems are also resampled with replacement, the fractions are 75\%, 53\% and 87\% (Figure~\ref{fig06:deficit}: 95\% interval $-0.86$ to $-0.23$).

\emph{Construction of the mid-infrared quantities.} $f_{\rm AGN}$ does not correlate with silicate depth ($\rho=-0.14$, $p=0.48$), although its component diagnostics do in opposite directions (Section~\ref{sec:discussion}), so the trend of Figure~\ref{fig06:deficit} is not inherited from the abscissa through $L_{\rm AGN,bol}$.

\begin{figure*}
    \centering
    \includegraphics[width=1\linewidth]{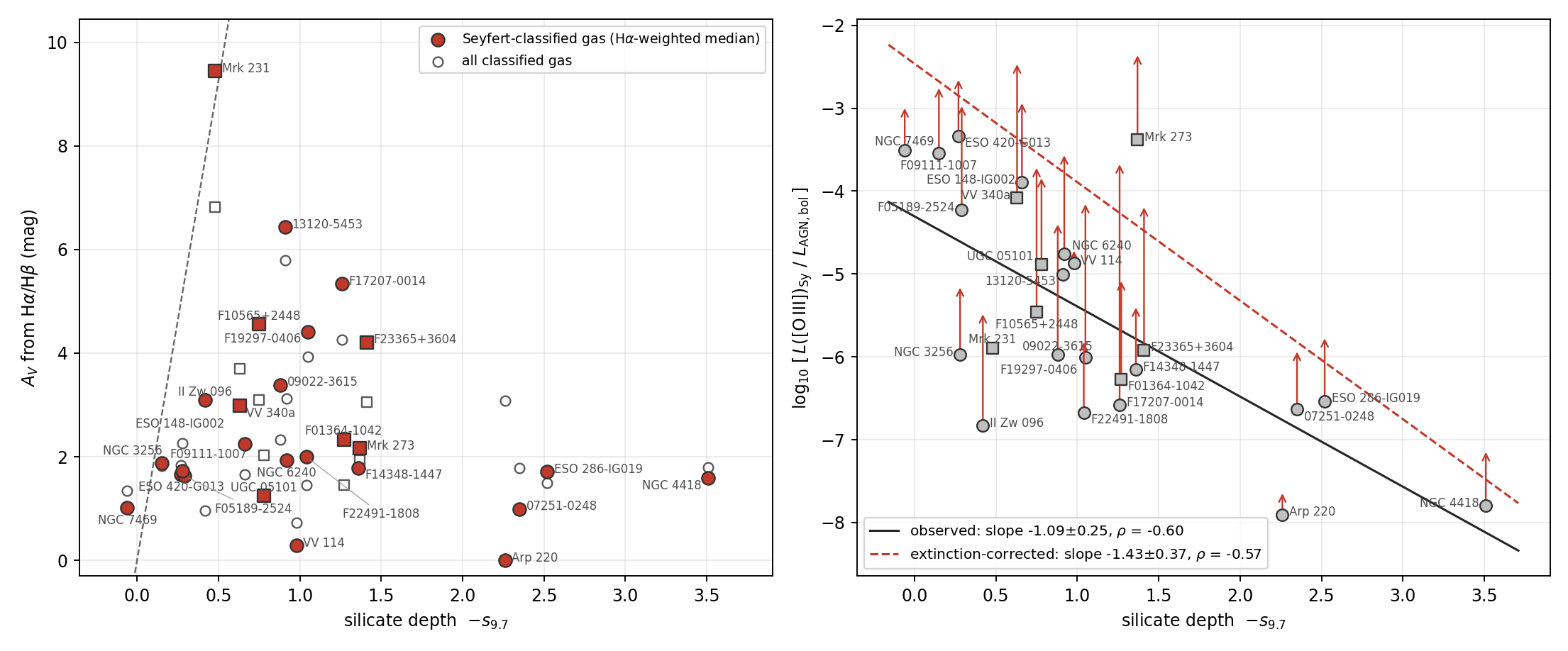}
    \caption{Foreground extinction does not explain Figure~\ref{fig06:deficit}. \emph{Left:} Balmer-decrement $A_V$ of the Seyfert-classified gas (filled; H$\alpha$-weighted median) and of all classified gas (open), against silicate depth; circles are MUSE and squares KCWI. The dashed line is the screen $A_V=18.5\,\tau_{9.7}$ implied by the silicate feature. \emph{Right:} $L(\oiii)_{\rm Sy}/L_{\rm AGN,bol}$ as observed (gray; solid fit) and after correcting each spaxel for its Balmer extinction (arrow tips; dashed fit).}
    \label{fig:app-balmer}
\end{figure*}

\emph{Foreground extinction.} Figure~\ref{fig:app-balmer} compares the Balmer-decrement extinction of the optical gas with the silicate depth. $A_V$ is computed spaxel by spaxel from H$\alpha$/H$\beta$ for an intrinsic ratio of 2.86 (Case B; \citealt{osterbrockferland06}) and the \cite{Cardelli89} curve with $R_V=3.1$, and summarized per system as the H$\alpha$-weighted median over the Seyfert-classified spaxels. It is uncorrelated with silicate depth ($\rho=-0.10$), with a median of 2.0 mag, an order of magnitude below the screen implied by the silicate feature, $A_V\approx18.5\,\tau_{9.7}$ \citep{Draine03} (median 17 mag, up to 65 mag). Correcting \oiii\ spaxel by spaxel leaves the trend of Figure~\ref{fig06:deficit} intact ($\rho=-0.57$) and steepens it (slope $-1.43\pm0.37$). The results are unchanged for an intrinsic ratio of 3.1, appropriate to AGN-ionized gas. The values for Mrk 231 ($A_V=9.4$) and possibly IRAS 13120$-$5453 (6.4) are inflated by the unresolved broad H$\alpha$ in their nuclei (Section~\ref{sec:fits}).

\begin{figure*}
    \centering
    \includegraphics[width=1\linewidth]{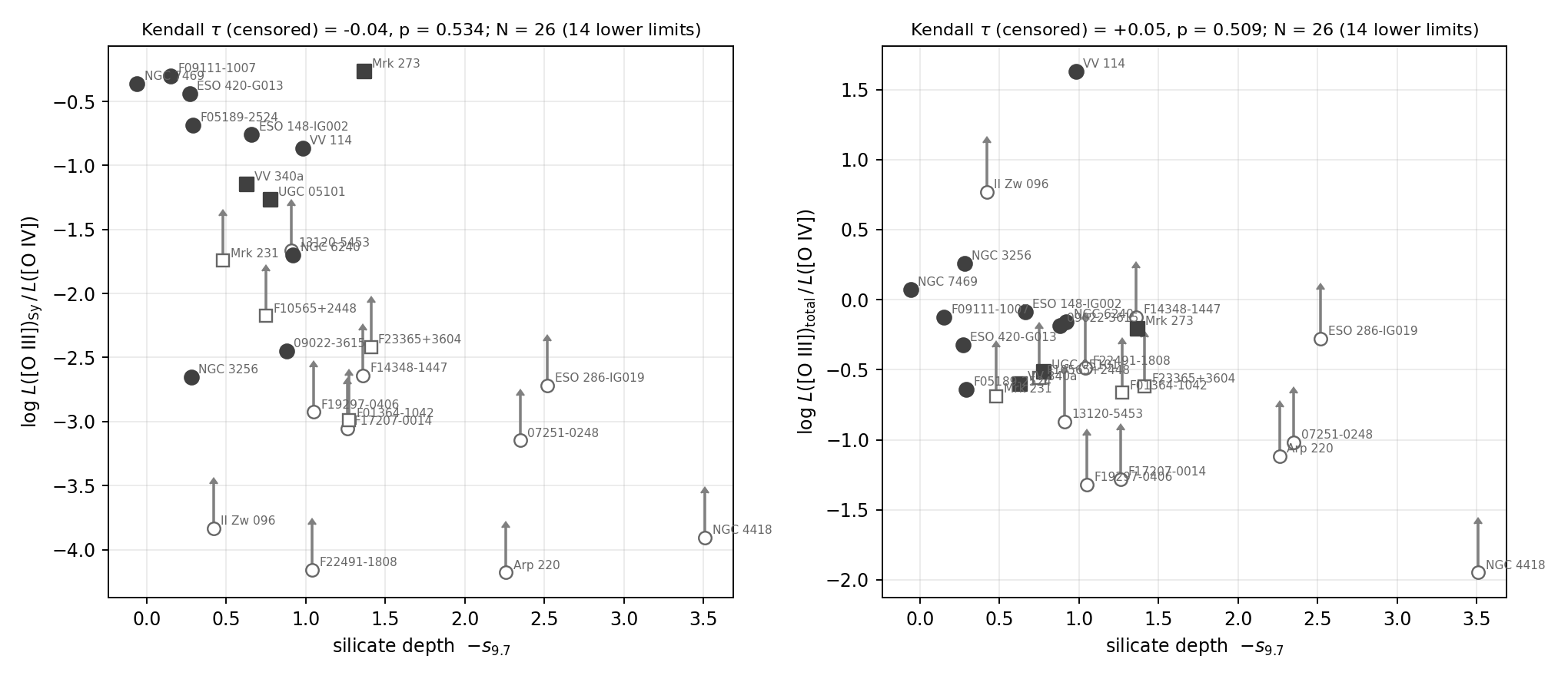}
    \caption{\oiii\ relative to the \emph{Spitzer}/IRS [O\,\textsc{iv}]\,25.89\,$\mu$m luminosity \citep{Inami13}, against silicate depth, for the Seyfert-classified gas (\emph{left}) and for all classified gas (\emph{right}). Open symbols with arrows are lower limits from [O\,\textsc{iv}] upper limits. The censored Kendall $\tau$ is given above each panel.}
    \label{fig:app-oiv}
\end{figure*}

\emph{An extinction-insensitive tracer.} Figure~\ref{fig:app-oiv} uses the \emph{Spitzer}/IRS [O\,\textsc{iv}]\,25.89\,$\mu$m fluxes of \cite{Inami13} as an AGN tracer that is nearly insensitive to dust. Neither $L(\oiii)_{\rm Sy}/L([{\rm O\,\textsc{iv}}])$ nor the ratio with \oiii\ summed over all classified gas shows a significant trend with depth (censored Kendall $\tau=-0.04$, $p=0.53$, and $\tau=+0.05$, $p=0.51$; \citealt{Isobe86}), but the test has no leverage where the effect is strongest: 14 of the 26 systems, including every one deeper than $-s_{9.7}\approx1.4$, have only an upper limit on [O\,\textsc{iv}]. The fact that the deepest nuclei are faint even in [O\,\textsc{iv}] is a strong case for analysis of the MIRI spectra in a future MAUNA paper.

\begin{figure}
    \centering
    \includegraphics[width=1\linewidth]{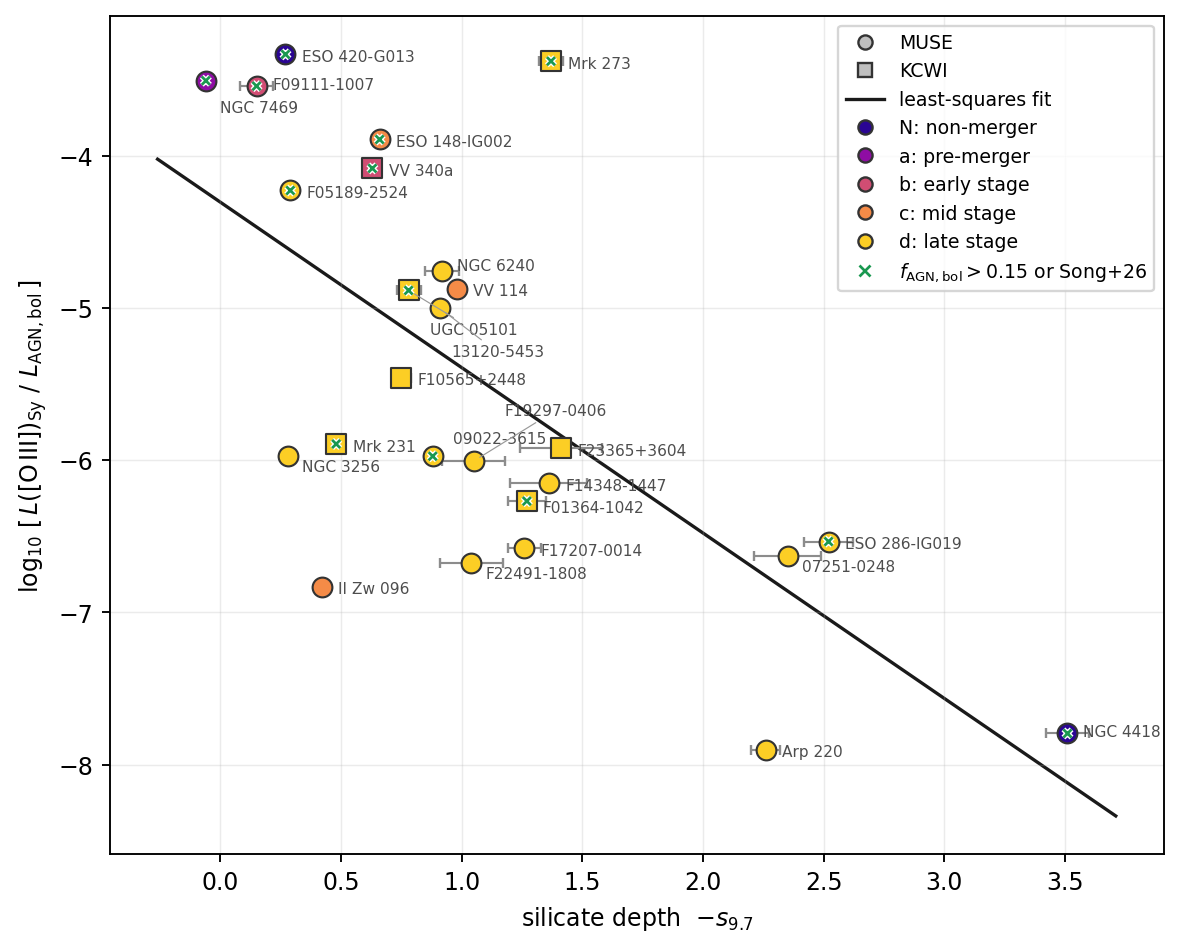}
    \caption{As Figure~\ref{fig06:deficit}, colored by the merger class of \cite{Stierwalt13}.}
    \label{fig:app-merger}
\end{figure}

\emph{Merger stage.} Figure~\ref{fig:app-merger} repeats Figure~\ref{fig06:deficit} colored by merger class. The residuals from the fit correlate with merger stage ($\rho=-0.50$, $p=0.0099$), with later stages lying lower, but 18 of the 26 systems are late-stage mergers, so stage and silicate depth cannot be separated in this sample.

\bibliography{references}{}

@ARTICLE{zhang18,
       author = {{Zhang}, Dong},
        title = "{A Review of the Theory of Galactic Winds Driven by Stellar Feedback}",
      journal = {Galaxies},
         year = 2018,
        month = nov,
       volume = {6},
       number = {4},
        pages = {114},
          doi = {10.3390/galaxies6040114},
archivePrefix = {arXiv},
       eprint = {1811.00558},
 primaryClass = {astro-ph.GA},
       adsurl = {https://ui.adsabs.harvard.edu/abs/2018Galax...6..114Z}
}

@ARTICLE{morganti17,
       author = {{Morganti}, Raffaella},
        title = "{The many routes to AGN feedback}",
      journal = {Frontiers in Astronomy and Space Sciences},
         year = 2017,
        month = nov,
       volume = {4},
          eid = {42},
        pages = {42},
          doi = {10.3389/fspas.2017.00042},
archivePrefix = {arXiv},
       eprint = {1712.05301},
 primaryClass = {astro-ph.GA},
       adsurl = {https://ui.adsabs.harvard.edu/abs/2017FrASS...4...42M}
}

@ARTICLE{dasyra16,
       author = {{Dasyra}, K.~M. and {Combes}, F. and {Oosterloo}, T. and {Oonk}, J.~B.~R. and {Morganti}, R. and {Salom{\'e}}, P. and {Vlahakis}, N.},
        title = "{ALMA reveals optically thin, highly excited CO gas in the jet-driven winds of the galaxy IC 5063}",
      journal = {\aap},
         year = 2016,
        month = nov,
       volume = {595},
          eid = {L7},
        pages = {L7},
          doi = {10.1051/0004-6361/201629689},
archivePrefix = {arXiv},
       eprint = {1609.03421},
 primaryClass = {astro-ph.GA},
       adsurl = {https://ui.adsabs.harvard.edu/abs/2016A&A...595L...7D}
}

@ARTICLE{emonts17,
       author = {{Emonts}, B.~H.~C. and {Colina}, L. and {Piqueras-L{\'o}pez}, J. and {Garcia-Burillo}, S. and {Pereira-Santaella}, M. and {Arribas}, S. and {Labiano}, A. and {Alonso-Herrero}, A.},
        title = "{Outflows of hot molecular gas in ultra-luminous infrared galaxies mapped with VLT-SINFONI}",
      journal = {\aap},
         year = 2017,
        month = nov,
       volume = {607},
          eid = {A116},
        pages = {A116},
          doi = {10.1051/0004-6361/201731508},
archivePrefix = {arXiv},
       eprint = {1708.09503},
 primaryClass = {astro-ph.GA},
       adsurl = {https://ui.adsabs.harvard.edu/abs/2017A&A...607A.116E}
}

@ARTICLE{barcos-munoz18,
       author = {{Barcos-Mu{\~n}oz}, Loreto and {Aalto}, Susanne and {Thompson}, Todd A. and {Sakamoto}, Kazushi and {Mart{\'\i}n}, Sergio and {Leroy}, Adam K. and {Privon}, George C. and {Evans}, Aaron S. and {Kepley}, Amanda},
        title = "{Fast, Collimated Outflow in the Western Nucleus of Arp 220}",
      journal = {\apjl},
         year = 2018,
        month = feb,
       volume = {853},
       number = {2},
          eid = {L28},
        pages = {L28},
          doi = {10.3847/2041-8213/aaa28d},
archivePrefix = {arXiv},
       eprint = {1712.06381},
 primaryClass = {astro-ph.GA},
       adsurl = {https://ui.adsabs.harvard.edu/abs/2018ApJ...853L..28B}
}

@ARTICLE{rodriguez-zaur13,
       author = {{Rodr{\'\i}guez Zaur{\'\i}n}, J. and {Tadhunter}, C.~N. and {Rose}, M. and {Holt}, J.},
        title = "{The importance of warm, AGN-driven outflows in the nuclear regions of nearby ULIRGs}",
      journal = {\mnras},
         year = 2013,
        month = jun,
       volume = {432},
       number = {1},
        pages = {138-166},
          doi = {10.1093/mnras/stt423},
archivePrefix = {arXiv},
       eprint = {1303.1400},
 primaryClass = {astro-ph.CO},
       adsurl = {https://ui.adsabs.harvard.edu/abs/2013MNRAS.432..138R}
}

@ARTICLE{pereira-santaella18,
       author = {{Pereira-Santaella}, M. and {Colina}, L. and {Garc{\'\i}a-Burillo}, S. and {Combes}, F. and {Emonts}, B. and {Aalto}, S. and {Alonso-Herrero}, A. and {Arribas}, S. and {Henkel}, C. and {Labiano}, A. and {Muller}, S. and {Piqueras L{\'o}pez}, J. and {Rigopoulou}, D. and {van der Werf}, P.},
        title = "{Spatially resolved cold molecular outflows in ULIRGs}",
      journal = {\aap},
         year = 2018,
        month = aug,
       volume = {616},
          eid = {A171},
        pages = {A171},
          doi = {10.1051/0004-6361/201833089},
archivePrefix = {arXiv},
       eprint = {1805.03667},
 primaryClass = {astro-ph.GA},
       adsurl = {https://ui.adsabs.harvard.edu/abs/2018A&A...616A.171P}
}

@ARTICLE{heckman17,
       author = {{Heckman}, Timothy M. and {Thompson}, Todd A.},
        title = "{Galactic Winds and the Role Played by Massive Stars}",
      journal = {arXiv e-prints},
         year = 2017,
        month = jan,
          eid = {arXiv:1701.09062},
        pages = {arXiv:1701.09062},
archivePrefix = {arXiv},
       eprint = {1701.09062},
 primaryClass = {astro-ph.GA},
       adsurl = {https://ui.adsabs.harvard.edu/abs/2017arXiv170109062H}
}

@ARTICLE{Kakkad18,
       author = {{Kakkad}, D. and {Groves}, B. and {Dopita}, M. and {Thomas}, A.~D. and {Davies}, R.~L. and {Mainieri}, V. and {Kharb}, P. and {Scharw{\"a}chter}, J. and {Hampton}, E.~J. and {Ho}, I. -T.},
        title = "{Spatially resolved electron density in the narrow line region of z < 0.02 radio AGNs}",
      journal = {\aap},
         year = 2018,
        month = oct,
       volume = {618},
          eid = {A6},
        pages = {A6},
          doi = {10.1051/0004-6361/201832790},
archivePrefix = {arXiv},
       eprint = {1806.02839},
 primaryClass = {astro-ph.GA},
       adsurl = {https://ui.adsabs.harvard.edu/abs/2018A&A...618A...6K}
}

@ARTICLE{Cicone18,
       author = {{Cicone}, Claudia and {Brusa}, Marcella and {Ramos Almeida}, Cristina and {Cresci}, Giovanni and {Husemann}, Bernd and {Mainieri}, Vincenzo},
        title = "{The largely unconstrained multiphase nature of outflows in AGN host galaxies}",
      journal = {Nature Astronomy},
         year = 2018,
        month = feb,
       volume = {2},
        pages = {176-178},
          doi = {10.1038/s41550-018-0406-3},
archivePrefix = {arXiv},
       eprint = {1802.10308},
 primaryClass = {astro-ph.GA},
       adsurl = {https://ui.adsabs.harvard.edu/abs/2018NatAs...2..176C}
}

@ARTICLE{Collins09,
       author = {{Collins}, N.~R. and {Kraemer}, S.~B. and {Crenshaw}, D.~M. and {Bruhweiler}, F.~C. and {Mel{\'e}ndez}, M.},
        title = "{Physical Conditions in the Narrow-Line Region of Markarian 3. II. Photoionization Modeling Results}",
      journal = {\apj},
         year = 2009,
        month = apr,
       volume = {694},
       number = {2},
        pages = {765-788},
          doi = {10.1088/0004-637X/694/2/765},
archivePrefix = {arXiv},
       eprint = {0901.0972},
 primaryClass = {astro-ph.GA},
       adsurl = {https://ui.adsabs.harvard.edu/abs/2009ApJ...694..765C}
}

@ARTICLE{Chen19,
       author = {{Chen}, Xiaoyang and {Akiyama}, Masayuki and {Noda}, Hirofumi and {Abdurro'uf}, Toba, Yoshiki and {Yamamura}, Issei and {Kawaguchi}, Toshihiro and {Kokubo}, Mitsuru and {Ichikawa}, Kohei},
        title = "{Discovery of a strong ionized-gas outflow in an AKARI-selected ultra-luminous infrared galaxy at z = 0.5}",
      journal = {\pasj},
         year = 2019,
        month = apr,
       volume = {71},
       number = {2},
          eid = {29},
        pages = {29},
          doi = {10.1093/pasj/psz002},
archivePrefix = {arXiv},
       eprint = {1901.05618},
 primaryClass = {astro-ph.GA},
       adsurl = {https://ui.adsabs.harvard.edu/abs/2019PASJ...71...29C}
}

@ARTICLE{Izumi20,
       author = {{Izumi}, Takuma and {Nguyen}, Dieu D. and {Imanishi}, Masatoshi and {Kawamuro}, Taiki and {Baba}, Shunsuke and {Nakano}, Suzuka and {Kohno}, Kotaro and {Matsushita}, Satoki and {Meier}, David S. and {Turner}, Jean L. and {Michiyama}, Tomonari and {Harada}, Nanase and {Mart{\'\i}n}, Sergio and {Nakanishi}, Kouichiro and {Takano}, Shuro and {Wiklind}, Tommy and {Nakai}, Naomasa and {Hsieh}, Pei-Ying},
        title = "{ALMA Observations of Multiple CO and C Lines toward the Active Galactic Nucleus of NGC 7469: An X-Ray-dominated Region Caught in the Act}",
      journal = {\apj},
         year = 2020,
        month = jul,
       volume = {898},
       number = {1},
          eid = {75},
        pages = {75},
          doi = {10.3847/1538-4357/ab9cb1},
archivePrefix = {arXiv},
       eprint = {2006.09406},
 primaryClass = {astro-ph.GA},
       adsurl = {https://ui.adsabs.harvard.edu/abs/2020ApJ...898...75I}
}

@ARTICLE{Leung21,
       author = {{Leung}, Gene C.~K. and {Coil}, Alison L. and {Rupke}, David S.~N. and {Perrotta}, Serena},
        title = "{KCWI Observations of the Extended Nebulae in Mrk 273}",
      journal = {\apj},
         year = 2021,
        month = jun,
       volume = {914},
       number = {1},
          eid = {17},
        pages = {17},
          doi = {10.3847/1538-4357/abf4da},
archivePrefix = {arXiv},
       eprint = {2011.09587},
 primaryClass = {astro-ph.GA},
       adsurl = {https://ui.adsabs.harvard.edu/abs/2021ApJ...914...17L}
}

@ARTICLE{Kauffmann03,
       author = {{Kauffmann}, Guinevere and {Heckman}, Timothy M. and {Tremonti}, Christy and {Brinchmann}, Jarle and {Charlot}, St{\'e}phane and {White}, Simon D.~M. and {Ridgway}, Susan E. and {Brinkmann}, Jon and {Fukugita}, Masataka and {Hall}, Patrick B. and {Ivezi{\'c}}, {\v{Z}}eljko and {Richards}, Gordon T. and {Schneider}, Donald P.},
        title = "{The host galaxies of active galactic nuclei}",
      journal = {\mnras},
         year = 2003,
        month = dec,
       volume = {346},
       number = {4},
        pages = {1055-1077},
          doi = {10.1111/j.1365-2966.2003.07154.x},
archivePrefix = {arXiv},
       eprint = {astro-ph/0304239},
 primaryClass = {astro-ph},
       adsurl = {https://ui.adsabs.harvard.edu/abs/2003MNRAS.346.1055K}
}

@ARTICLE{Cappellari17,
       author = {{Cappellari}, Michele},
        title = "{Improving the full spectrum fitting method: accurate convolution with Gauss-Hermite functions}",
      journal = {\mnras},
         year = 2017,
        month = apr,
       volume = {466},
       number = {1},
        pages = {798-811},
          doi = {10.1093/mnras/stw3020},
archivePrefix = {arXiv},
       eprint = {1607.08538},
 primaryClass = {astro-ph.GA},
       adsurl = {https://ui.adsabs.harvard.edu/abs/2017MNRAS.466..798C}
}

@ARTICLE{GarciaBurillo15,
       author = {{Garc{\'\i}a-Burillo}, S. and {Combes}, F. and {Usero}, A. and
         {Aalto}, S. and {Colina}, L. and {Alonso-Herrero}, A. and
         {Hunt}, L.~K. and {Arribas}, S. and {Costagliola}, F. and
         {Labiano}, A. and {Neri}, R. and {Pereira-Santaella}, M. and
         {Tacconi}, L.~J. and {van der Werf}, P.~P.},
        title = "{High-resolution imaging of the molecular outflows in two mergers: <ASTROBJ>IRAS 17208-0014</ASTROBJ> and <ASTROBJ>NGC 1614</ASTROBJ>}",
      journal = {\aap},
         year = 2015,
        month = aug,
       volume = {580},
          eid = {A35},
        pages = {A35},
          doi = {10.1051/0004-6361/201526133},
archivePrefix = {arXiv},
       eprint = {1505.04705},
 primaryClass = {astro-ph.GA},
       adsurl = {https://ui.adsabs.harvard.edu/abs/2015A&A...580A..35G}
}

@article{Cardelli89,
	Adsurl = {http://adsabs.harvard.edu/abs/1989ApJ...345..245C},
	Author = {{Cardelli}, J.~A. and {Clayton}, G.~C. and {Mathis}, J.~S.},
	Doi = {10.1086/167900},
	Journal = {\apj},
	Month = oct,
	Pages = {245-256},
	Title = {{The relationship between infrared, optical, and ultraviolet extinction}},
	Volume = 345,
	Year = 1989}

@article{Diaz-Santos17,
	Adsurl = {http://adsabs.harvard.edu/abs/2017ApJ...846...32D},
	Archiveprefix = {arXiv},
	Author = {{D{\'{\i}}az-Santos}, T. and {Armus}, L. and {Charmandaris}, V. and {Lu}, N. and {Stierwalt}, S. and {Stacey}, G. and {Malhotra}, S. and {van der Werf}, P.~P. and {Howell}, J.~H. and {Privon}, G.~C. and {Mazzarella}, J.~M. and {Goldsmith}, P.~F. and {Murphy}, E.~J. and {Barcos-Mu{\~n}oz}, L. and {Linden}, S.~T. and {Inami}, H. and {Larson}, K.~L. and {Evans}, A.~S. and {Appleton}, P. and {Iwasawa}, K. and {Lord}, S. and {Sanders}, D.~B. and {Surace}, J.~A.},
	Doi = {10.3847/1538-4357/aa81d7},
	Eid = {32},
	Eprint = {1705.04326},
	Journal = {\apj},
	Month = sep,
	Pages = {32},
	Title = {{A Herschel/PACS Far-infrared Line Emission Survey of Local Luminous Infrared Galaxies}},
	Volume = 846,
	Year = 2017}

@article{Inami13,
	Adsurl = {http://adsabs.harvard.edu/abs/2013ApJ...777..156I},
	Archiveprefix = {arXiv},
	Author = {{Inami}, H. and {Armus}, L. and {Charmandaris}, V. and {Groves}, B. and {Kewley}, L. and {Petric}, A. and {Stierwalt}, S. and {D{\'{\i}}az-Santos}, T. and {Surace}, J. and {Rich}, J. and {Haan}, S. and {Howell}, J. and {Evans}, A.~S. and {Mazzarella}, J. and {Marshall}, J. and {Appleton}, P. and {Lord}, S. and {Spoon}, H. and {Frayer}, D. and {Matsuhara}, H. and {Veilleux}, S.},
	Doi = {10.1088/0004-637X/777/2/156},
	Eid = {156},
	Eprint = {1309.4788},
	Journal = {\apj},
	Month = nov,
	Pages = {156},
	Title = {{Mid-infrared Atomic Fine-structure Emission-line Spectra of Luminous Infrared Galaxies: Spitzer/IRS Spectra of the GOALS Sample}},
	Volume = 777,
	Year = 2013}

@article{Kewley01,
	Adsurl = {http://adsabs.harvard.edu/abs/2001ApJ...556..121K},
	Author = {{Kewley}, L.~J. and {Dopita}, M.~A. and {Sutherland}, R.~S. and {Heisler}, C.~A. and {Trevena}, J.},
	Doi = {10.1086/321545},
	Eprint = {astro-ph/0106324},
	Journal = {\apj},
	Month = jul,
	Pages = {121-140},
	Title = {{Theoretical Modeling of Starburst Galaxies}},
	Volume = 556,
	Year = 2001}

@article{Rich15,
	Adsurl = {http://adsabs.harvard.edu/abs/2015ApJS..221...28R},
	Archiveprefix = {arXiv},
	Author = {{Rich}, J.~A. and {Kewley}, L.~J. and {Dopita}, M.~A.},
	Doi = {10.1088/0067-0049/221/2/28},
	Eid = {28},
	Eprint = {1509.08468},
	Journal = {\apjs},
	Month = dec,
	Pages = {28},
	Title = {{Galaxy Mergers Drive Shocks: An Integral Field Study of GOALS Galaxies}},
	Volume = 221,
	Year = 2015}

@article{Sanders03,
	Adsurl = {http://adsabs.harvard.edu/abs/2003AJ....126.1607S},
	Author = {{Sanders}, D.~B. and {Mazzarella}, J.~M. and {Kim}, D.-C. and {Surace}, J.~A. and {Soifer}, B.~T.},
	Doi = {10.1086/376841},
	Eprint = {astro-ph/0306263},
	Journal = {\aj},
	Month = oct,
	Pages = {1607-1664},
	Title = {{The IRAS Revised Bright Galaxy Sample}},
	Volume = 126,
	Year = 2003}

@article{Soto12,
	Adsurl = {http://adsabs.harvard.edu/abs/2012ApJ...757...86S},
	Archiveprefix = {arXiv},
	Author = {{Soto}, K.~T. and {Martin}, C.~L. and {Prescott}, M.~K.~M. and {Armus}, L.},
	Doi = {10.1088/0004-637X/757/1/86},
	Eid = {86},
	Eprint = {1205.0083},
	Journal = {ApJ},
	Month = sep,
	Pages = {86},
	Title = {{The Emission-line Spectra of Major Mergers: Evidence for Shocked Outflows}},
	Volume = 757,
	Year = 2012}

@article{Spoon07,
	Adsurl = {http://adsabs.harvard.edu/abs/2007ApJ...654L..49S},
	Author = {{Spoon}, H.~W.~W. and {Marshall}, J.~A. and {Houck}, J.~R. and {Elitzur}, M. and {Hao}, L. and {Armus}, L. and {Brandl}, B.~R. and {Charmandaris}, V.},
	Doi = {10.1086/511268},
	Eprint = {astro-ph/0611918},
	Journal = {\apjl},
	Month = jan,
	Pages = {L49-L52},
	Title = {{Mid-Infrared Galaxy Classification Based on Silicate Obscuration and PAH Equivalent Width}},
	Volume = 654,
	Year = 2007}

@article{Stierwalt14,
	Adsurl = {http://adsabs.harvard.edu/abs/2014ApJ...790..124S},
	Archiveprefix = {arXiv},
	Author = {{Stierwalt}, S. and {Armus}, L. and {Charmandaris}, V. and {Diaz-Santos}, T. and {Marshall}, J. and {Evans}, A.~S. and {Haan}, S. and {Howell}, J. and {Iwasawa}, K. and {Kim}, D.~C. and {Murphy}, E.~J. and {Rich}, J.~A. and {Spoon}, H.~W.~W. and {Inami}, H. and {Petric}, A.~O. and {U}, V.},
	Doi = {10.1088/0004-637X/790/2/124},
	Eid = {124},
	Eprint = {1406.3891},
	Journal = {\apj},
	Month = aug,
	Pages = {124},
	Title = {{Mid-infrared Properties of Luminous Infrared Galaxies. II. Probing the Dust and Gas Physics of the GOALS Sample}},
	Volume = 790,
	Year = 2014}

@article{U19,
	Adsurl = {http://adsabs.harvard.edu/abs/2019ApJ...871..166U},
	Author = {{U}, V. and {Medling}, A.~M. and {Inami}, H. and {Armus}, L. and {D{\'{\i}}az-Santos}, T. and {Charmandaris}, V. and {Howell}, J. and {Stierwalt}, S. and {Privon}, G.~C. and {Linden}, S.~T. and {Sanders}, D.~B. and {Max}, C.~E. and {Evans}, A.~S. and {Barcos-Mu{\~n}oz}, L. and {Chiang}, C.~W.~K. and {Appleton}, P. and {Canalizo}, G. and {Fazio}, G. and {Iwasawa}, K. and {Larson}, K. and {Mazzarella}, J. and {Murphy}, E. and {Rich}, J. and {Surace}, J.},
	Doi = {10.3847/1538-4357/aaf1c2},
	Eid = {166},
	Journal = {\apj},
	Month = feb,
	Pages = {166},
	Title = {{Keck OSIRIS AO LIRG Analysis (KOALA): Feedback in the Nuclei of Luminous Infrared Galaxies}},
	Volume = 871,
	Year = 2019}

@article{Veilleux87,
	Adsurl = {http://adsabs.harvard.edu/abs/1987ApJS...63..295V},
	Author = {{Veilleux}, S. and {Osterbrock}, D.~E.},
	Doi = {10.1086/191166},
	Journal = {ApJS},
	Month = feb,
	Pages = {295-310},
	Title = {{Spectral classification of emission-line galaxies}},
	Volume = 63,
	Year = 1987}

@article{Yuan10,
	Adsurl = {http://adsabs.harvard.edu/abs/2010ApJ...709..884Y},
	Archiveprefix = {arXiv},
	Author = {{Yuan}, T.-T. and {Kewley}, L.~J. and {Sanders}, D.~B.},
	Doi = {10.1088/0004-637X/709/2/884},
	Eprint = {0911.3728},
	Journal = {\apj},
	Month = feb,
	Pages = {884-911},
	Title = {{The Role of Starburst-Active Galactic Nucleus Composites in Luminous Infrared Galaxy Mergers: Insights from the New Optical Classification Scheme}},
	Volume = 709,
	Year = 2010}

@article{Veilleux05,
	Adsurl = {http://adsabs.harvard.edu/abs/2005ARA%26A..43..769V},
	Author = {{Veilleux}, S. and {Cecil}, G. and {Bland-Hawthorn}, J.},
	Doi = {10.1146/annurev.astro.43.072103.150610},
	Eprint = {astro-ph/0504435},
	Journal = {\araa},
	Month = sep,
	Pages = {769-826},
	Title = {{Galactic Winds}},
	Volume = 43,
	Year = 2005}

@article{Diaz-Santos11,
	Adsurl = {http://adsabs.harvard.edu/abs/2011ApJ...741...32D},
	Archiveprefix = {arXiv},
	Author = {{D{\'{\i}}az-Santos}, T. and {Charmandaris}, V. and {Armus}, L. and {Stierwalt}, S. and {Haan}, S. and {Mazzarella}, J.~M. and {Howell}, J.~H. and {Veilleux}, S. and {Murphy}, E.~J. and {Petric}, A.~O. and {Appleton}, P. and {Evans}, A.~S. and {Sanders}, D.~B. and {Surace}, J.~A.},
	Doi = {10.1088/0004-637X/741/1/32},
	Eid = {32},
	Eprint = {1107.5958},
	Journal = {ApJ},
	Month = nov,
	Pages = {32},
	Title = {{The Spatial Extent of (U)LIRGS in the Mid-infrared. II. Feature Emission}},
	Volume = 741,
	Year = 2011}

@article{Baldwin81,
	Adsurl = {http://adsabs.harvard.edu/abs/1981PASP...93....5B},
	Author = {{Baldwin}, J.~A. and {Phillips}, M.~M. and {Terlevich}, R.},
	Doi = {10.1086/130766},
	Journal = {PASP},
	Month = feb,
	Pages = {5-19},
	Title = {{Classification parameters for the emission-line spectra of extragalactic objects}},
	Volume = 93,
	Year = 1981}

@article{Kewley06,
	Adsurl = {http://adsabs.harvard.edu/abs/2006MNRAS.372..961K},
	Author = {{Kewley}, L.~J. and {Groves}, B. and {Kauffmann}, G. and {Heckman}, T.},
	Doi = {10.1111/j.1365-2966.2006.10859.x},
	Eprint = {astro-ph/0605681},
	Journal = {MNRAS},
	Month = nov,
	Pages = {961-976},
	Title = {{The host galaxies and classification of active galactic nuclei}},
	Volume = 372,
	Year = 2006}

@article{Diaz-Santos10,
	Adsurl = {http://adsabs.harvard.edu/abs/2010ApJ...723..993D},
	Archiveprefix = {arXiv},
	Author = {{D{\'{\i}}az-Santos}, T. and {Charmandaris}, V. and {Armus}, L. and {Petric}, A.~O. and {Howell}, J.~H. and {Murphy}, E.~J. and {Mazzarella}, J.~M. and {Veilleux}, S. and {Bothun}, G. and {Inami}, H. and {Appleton}, P.~N. and {Evans}, A.~S. and {Haan}, S. and {Marshall}, J.~A. and {Sanders}, D.~B. and {Stierwalt}, S. and {Surace}, J.~A.},
	Doi = {10.1088/0004-637X/723/2/993},
	Eprint = {1009.0038},
	Journal = {ApJ},
	Month = nov,
	Pages = {993-1005},
	Primaryclass = {astro-ph.CO},
	Title = {{The Spatial Extent of (U)LIRGs in the Mid-infrared. I. The Continuum Emission}},
	Volume = 723,
	Year = 2010}

@article{Stierwalt13,
	Adsurl = {http://adsabs.harvard.edu/abs/2013ApJS..206....1S},
	Archiveprefix = {arXiv},
	Author = {{Stierwalt}, S. and {Armus}, L. and {Surace}, J.~A. and {Inami}, H. and {Petric}, A.~O. and {Diaz-Santos}, T. and {Haan}, S. and {Charmandaris}, V. and {Howell}, J. and {Kim}, D.~C. and {Marshall}, J. and {Mazzarella}, J.~M. and {Spoon}, H.~W.~W. and {Veilleux}, S. and {Evans}, A. and {Sanders}, D.~B. and {Appleton}, P. and {Bothun}, G. and {Bridge}, C.~R. and {Chan}, B. and {Frayer}, D. and {Iwasawa}, K. and {Kewley}, L.~J. and {Lord}, S. and {Madore}, B.~F. and {Melbourne}, J.~E. and {Murphy}, E.~J. and {Rich}, J.~A. and {Schulz}, B. and {Sturm}, E. and {Vavilkin}, T. and {Xu}, K.},
	Doi = {10.1088/0067-0049/206/1/1},
	Eid = {1},
	Eprint = {1302.4477},
	Journal = {ApJS},
	Month = may,
	Pages = {1},
	Primaryclass = {astro-ph.CO},
	Title = {{Mid-infrared Properties of Nearby Luminous Infrared Galaxies. I. Spitzer Infrared Spectrograph Spectra for the GOALS Sample}},
	Volume = 206,
	Year = 2013}

@article{Rich11,
	Adsurl = {http://adsabs.harvard.edu/abs/2011ApJ...734...87R},
	Archiveprefix = {arXiv},
	Author = {{Rich}, J.~A. and {Kewley}, L.~J. and {Dopita}, M.~A.},
	Doi = {10.1088/0004-637X/734/2/87},
	Eid = {87},
	Eprint = {1104.1177},
	Journal = {ApJ},
	Month = jun,
	Pages = {87},
	Primaryclass = {astro-ph.CO},
	Title = {{Galaxy-wide Shocks in Late-merger Stage Luminous Infrared Galaxies}},
	Volume = 734,
	Year = 2011}

@article{Armus09,
	Adsurl = {http://adsabs.harvard.edu/abs/2009PASP..121..559A},
	Archiveprefix = {arXiv},
	Author = {{Armus}, L. and {Mazzarella}, J.~M. and {Evans}, A.~S. and {Surace}, J.~A. and {Sanders}, D.~B. and {Iwasawa}, K. and {Frayer}, D.~T. and {Howell}, J.~H. and {Chan}, B. and {Petric}, A. and {Vavilkin}, T. and {Kim}, D.~C. and {Haan}, S. and {Inami}, H. and {Murphy}, E.~J. and {Appleton}, P.~N. and {Barnes}, J.~E. and {Bothun}, G. and {Bridge}, C.~R. and {Charmandaris}, V. and {Jensen}, J.~B. and {Kewley}, L.~J. and {Lord}, S. and {Madore}, B.~F. and {Marshall}, J.~A. and {Melbourne}, J.~E. and {Rich}, J. and {Satyapal}, S. and {Schulz}, B. and {Spoon}, H.~W.~W. and {Sturm}, E. and {U}, V. and {Veilleux}, S. and {Xu}, K.},
	Doi = {10.1086/600092},
	Eprint = {0904.4498},
	Journal = {PASP},
	Month = jun,
	Pages = {559-576},
	Primaryclass = {astro-ph.CO},
	Title = {{GOALS: The Great Observatories All-Sky LIRG Survey}},
	Volume = 121,
	Year = 2009}

@article{U13,
	Adsurl = {http://adsabs.harvard.edu/abs/2013ApJ...775..115U},
	Archiveprefix = {arXiv},
	Author = {{U}, V. and {Medling}, A. and {Sanders}, D. and {Max}, C. and {Armus}, L. and {Iwasawa}, K. and {Evans}, A. and {Kewley}, L. and {Fazio}, G.},
	Doi = {10.1088/0004-637X/775/2/115},
	Eid = {115},
	Eprint = {1307.8440},
	Journal = {ApJ},
	Month = oct,
	Pages = {115},
	Primaryclass = {astro-ph.CO},
	Title = {{The Inner Kiloparsec of Mrk 273 with Keck Adaptive Optics}},
	Volume = 775,
	Year = 2013}

@ARTICLE{morrisey18,
       author = {{Morrissey}, Patrick and {Matuszewski}, Matuesz and {Martin}, D. Christopher and {Neill}, James D. and {Epps}, Harland and {Fucik}, Jason and {Weber}, Bob and {Darvish}, Behnam and {Adkins}, Sean and {Allen}, Steve and {Bartos}, Randy and {Belicki}, Justin and {Cabak}, Jerry and {Callahan}, Shawn and {Cowley}, Dave and {Crabill}, Marty and {Deich}, Willian and {Delecroix}, Alex and {Doppman}, Greg and {Hilyard}, David and {James}, Ean and {Kaye}, Steve and {Kokorowski}, Michael and {Kwok}, Shui and {Lanclos}, Kyle and {Milner}, Steve and {Moore}, Anna and {O'Sullivan}, Donal and {Parihar}, Prachi and {Park}, Sam and {Phillips}, Andrew and {Rizzi}, Luca and {Rockosi}, Constance and {Rodriguez}, Hector and {Salaun}, Yves and {Seaman}, Kirk and {Sheikh}, David and {Weiss}, Jason and {Zarzaca}, Ray},
        title = "{The Keck Cosmic Web Imager Integral Field Spectrograph}",
      journal = {\apj},
         year = 2018,
        month = sep,
       volume = {864},
       number = {1},
          eid = {93},
        pages = {93},
          doi = {10.3847/1538-4357/aad597},
archivePrefix = {arXiv},
       eprint = {1807.10356},
 primaryClass = {astro-ph.IM},
       adsurl = {https://ui.adsabs.harvard.edu/abs/2018ApJ...864...93M}
}

@ARTICLE{osullivan20,
       author = {{O'Sullivan}, Donal and {Chen}, Yuguang},
        title = "{CWITools: A Python3 Data Analysis Pipeline for the Cosmic Web Imager Instruments}",
      journal = {arXiv e-prints},
         year = 2020,
        month = nov,
          eid = {arXiv:2011.05444},
        pages = {arXiv:2011.05444},
          doi = {10.48550/arXiv.2011.05444},
archivePrefix = {arXiv},
       eprint = {2011.05444},
 primaryClass = {astro-ph.IM},
       adsurl = {https://ui.adsabs.harvard.edu/abs/2020arXiv201105444O}
}

@BOOK{osterbrockferland06,
       author = {{Osterbrock}, Donald E. and {Ferland}, Gary J.},
        title = "{Astrophysics of gaseous nebulae and active galactic nuclei}",
         year = 2006,
       adsurl = {https://ui.adsabs.harvard.edu/abs/2006agna.book.....O}
}

@ARTICLE{bacon17,
       author = {{Bacon}, Roland and {Conseil}, Simon and {Mary}, David and {Brinchmann}, Jarle and {Shepherd}, Martin and {Akhlaghi}, Mohammad and {Weilbacher}, Peter M. and {Piqueras}, Laure and {Wisotzki}, Lutz and {Lagattuta}, David and {Epinat}, Benoit and {Guerou}, Adrien and {Inami}, Hanae and {Cantalupo}, Sebastiano and {Courbot}, Jean Baptiste and {Contini}, Thierry and {Richard}, Johan and {Maseda}, Michael and {Bouwens}, Rychard and {Bouch{\'e}}, Nicolas and {Kollatschny}, Wolfram and {Schaye}, Joop and {Marino}, Raffaella Anna and {Pello}, Roser and {Herenz}, Christian and {Guiderdoni}, Bruno and {Carollo}, Marcella},
        title = "{The MUSE Hubble Ultra Deep Field Survey. I. Survey description, data reduction, and source detection}",
      journal = {\aap},
         year = 2017,
        month = dec,
       volume = {608},
          eid = {A1},
        pages = {A1},
          doi = {10.1051/0004-6361/201730833},
archivePrefix = {arXiv},
       eprint = {1710.03002},
 primaryClass = {astro-ph.GA},
       adsurl = {https://ui.adsabs.harvard.edu/abs/2017A&A...608A...1B}
}

@ARTICLE{schawinski07,
       author = {{Schawinski}, Kevin and {Thomas}, Daniel and {Sarzi}, Marc and {Maraston}, Claudia and {Kaviraj}, Sugata and {Joo}, Seok-Joo and {Yi}, Sukyoung K. and {Silk}, Joseph},
        title = "{Observational evidence for AGN feedback in early-type galaxies}",
      journal = {\mnras},
         year = 2007,
        month = dec,
       volume = {382},
       number = {4},
        pages = {1415-1431},
          doi = {10.1111/j.1365-2966.2007.12487.x},
archivePrefix = {arXiv},
       eprint = {0709.3015},
 primaryClass = {astro-ph},
       adsurl = {https://ui.adsabs.harvard.edu/abs/2007MNRAS.382.1415S}
}

@ARTICLE{veilleux09,
       author = {{Veilleux}, S. and {Rupke}, D.~S.~N. and {Kim}, D.-C. and {Genzel}, R. and {Sturm}, E. and {Lutz}, D. and {Contursi}, A. and {Schweitzer}, M. and {Tacconi}, L.~J. and {Netzer}, H. and {Sternberg}, A. and {Mihos}, J.~C. and {Baker}, A.~J. and {Mazzarella}, J.~M. and {Lord}, S. and {Sanders}, D.~B. and {Stockton}, A. and {Joseph}, R.~D. and {Barnes}, J.~E.},
        title = "{Spitzer Quasar and Ulirg Evolution Study (QUEST). IV. Comparison of 1 Jy Ultraluminous Infrared Galaxies with Palomar-Green Quasars}",
      journal = {\apjs},
         year = 2009,
        month = jun,
       volume = {182},
       number = {2},
        pages = {628-666},
          doi = {10.1088/0067-0049/182/2/628},
archivePrefix = {arXiv},
       eprint = {0905.1577},
 primaryClass = {astro-ph.CO},
       adsurl = {https://ui.adsabs.harvard.edu/abs/2009ApJS..182..628V}
}

@ARTICLE{Goulding09,
   author = {{Goulding}, A.~D. and {Alexander}, D.~M.},
    title = "{Towards a complete census of AGN in nearby Galaxies: a large population of optically unidentified AGN}",
  journal = {\mnras},
     year = 2009,
   volume = 398,
    pages = {1165-1193},
      doi = {10.1111/j.1365-2966.2009.15194.x}
}

@ARTICLE{Bennert02,
   author = {{Bennert}, N. and {Falcke}, H. and {Schulz}, H. and {Wilson}, A.~S. and {Wills}, B.~J.},
    title = "{Size and Structure of the Narrow-Line Region of Quasars}",
  journal = {\apjl},
     year = 2002,
   volume = 574,
    pages = {L105-L109},
      doi = {10.1086/342420}
}

@ARTICLE{Greene11,
   author = {{Greene}, J.~E. and {Zakamska}, N.~L. and {Ho}, L.~C. and {Barth}, A.~J.},
    title = "{Feedback in Luminous Obscured Quasars}",
  journal = {\apj},
     year = 2011,
   volume = 732,
    pages = {9},
      doi = {10.1088/0004-637X/732/1/9}
}

@ARTICLE{Hainline13,
   author = {{Hainline}, K.~N. and {Hickox}, R. and {Greene}, J.~E. and {Myers}, A.~D. and {Zakamska}, N.~L.},
    title = "{Gemini Long-slit Observations of Luminous Obscured Quasars: Further Evidence for an Upper Limit on the Size of the Narrow-line Region}",
  journal = {\apj},
     year = 2013,
   volume = 774,
    pages = {145},
      doi = {10.1088/0004-637X/774/2/145}
}

@ARTICLE{Draine03,
   author = {{Draine}, B.~T.},
    title = "{Interstellar Dust Grains}",
  journal = {\araa},
     year = 2003,
   volume = 41,
    pages = {241-289},
      doi = {10.1146/annurev.astro.41.011802.094840}
}

@ARTICLE{Lupton04,
   author = {{Lupton}, R. and {Blanton}, M.~R. and {Fekete}, G. and {Hogg}, D.~W. and {O'Mullane}, W. and {Szalay}, A. and {Wherry}, N.},
    title = "{Preparing Red-Green-Blue Images from CCD Data}",
  journal = {\pasp},
     year = 2004,
   volume = 116,
    pages = {133-137},
      doi = {10.1086/382245}
}

@ARTICLE{Vazdekis16,
   author = {{Vazdekis}, A. and {Koleva}, M. and {Ricciardelli}, E. and {R{\"o}ck}, B. and {Falc{\'o}n-Barroso}, J.},
    title = "{UV-extended E-MILES stellar population models: young components in massive early-type galaxies}",
  journal = {\mnras},
     year = 2016,
   volume = 463,
    pages = {3409-3436},
      doi = {10.1093/mnras/stw2231}
}

@ARTICLE{Isobe86,
   author = {{Isobe}, T. and {Feigelson}, E.~D. and {Nelson}, P.~I.},
    title = "{Statistical Methods for Astronomical Data with Upper Limits. II. Correlation and Regression}",
  journal = {\apj},
     year = 1986,
   volume = 306,
    pages = {490},
      doi = {10.1086/164359}
}

@ARTICLE{astropy22,
   author = {{Astropy Collaboration}},
    title = "{The Astropy Project: Sustaining and Growing a Community-oriented Open-source Project and the Latest Major Release (v5.0) of the Core Package}",
  journal = {\apj},
     year = 2022,
   volume = 935,
    pages = {167},
      doi = {10.3847/1538-4357/ac7c74}
}

@misc{Kader26_HyperCube,
   author = {{Kader}, Justin},
    title = "{HyperCube}",
     year = 2026,
    month = oct,
  version = {v0.5.1},
howpublished = {Zenodo},
      doi = {10.5281/zenodo.23247289},
      url = {https://doi.org/10.5281/zenodo.23247289}
}

@article{glenn25,
author = {Jason Glenn and Margaret Meixner and Charles  M. Bradford and Klaus Pontoppidan and Alexandra Pope and Tiffany Kataria and Jennifer Rocca and Elizabeth Luthman and Lee Armus and Jochem Baselmans and Cara Battersby and Alberto Bolatto and Denis Burgarella and Weibo Chen and Laure Ciesla and Peter Day and Anna Di Giorgio and Michael DiPirro and Charles Darren Dowell and Pierre Echternach and Thomas Essinger-Hileman and Marc Foote and Carlotta Gruppioni and Brandon Hensley and Thomas Henning and Willem Jellema and Matthew Johnson and Alan Kogut and Oliver Krause and James McGuire and Elisabeth Mills and Arielle Moullet and Michael Rodgers and Marc Sauvage and John  D. Smith and Rachel Somerville and Johannes Staguhn and Thomas Stevenson and Carole Tucker and Stephen Unwin and John Ziemer and Matthew Cannella and Richard Dissly},
title = {{PRIMA mission concept}},
volume = {11},
journal = {Journal of Astronomical Telescopes, Instruments, and Systems},
number = {3},
publisher = {SPIE},
pages = {031628},
year = {2025},
doi = {10.1117/1.JATIS.11.3.031628},
URL = {https://doi.org/10.1117/1.JATIS.11.3.031628}
}

@ARTICLE{song26,
       author = {{Song}, Y. and {U}, V. and {Kader}, J. and {Bianchin}, M. and {Agostino}, J. and {Barcos-Mu{\~n}oz}, L. and {Torres-Alb{\`a}}, N. and {Medling}, A. and {Ricci}, C. and {Armus}, L. and {Lai}, T.~S.-Y. and {Ghodsi}, L. and {Evans}, A. and {Inami}, H. and {Stierwalt}, S. and {Bohn}, T. and {Emig}, K.~L. and {Buiten}, V. and {Donnelly}, G. and {Treister}, E. and {Jimenez-Gallardo}, A. and {B{\"o}ker}, T. and {Kakkad}, D. and {Linden}, S.~T. and {van der Werf}, P. and {Charmandaris}, V. and {Remigio}, R. and {Lenki{\'c}}, L. and {Finlez}, C. and {S{\'a}nchez-Garc{\'\i}a}, M. and {Lofaro}, C. and {Saravia}, A. and {Yoon}, I. and {Privon}, G.~C. and {Gao}, T. and {Aziz}, H. and {McGurk}, R. and {D{\'\i}az-Santos}, T. and {Kunneriath}, D. and {Sanders}, David B.},
        title = "{The multiphase biconical outflow in the local infrared-luminous merger IRAS F01364-1042}",
      journal = {\aap},
         year = 2026,
        month = aug,
       volume = {712},
          eid = {A59},
        pages = {A59},
          doi = {10.1051/0004-6361/202557994},
archivePrefix = {arXiv},
       eprint = {2606.06005},
 primaryClass = {astro-ph.GA},
       adsurl = {https://ui.adsabs.harvard.edu/abs/2026A&A...712A..59S}
}

@ARTICLE{kader26,
       author = {{Kader}, Justin A. and {U}, Vivian and {Barcos-Mu{\~n}oz}, Loreto and {Bianchin}, Marina and {Song}, Yiqing and {Linden}, Sean T. and {Canalizo}, Gabriela and {Aravindan}, Archana and {Privon}, George C. and {D{\'\i}az-Santos}, Tanio and {Hayward}, Christopher and {Malkan}, Matthew A. and {Armus}, Lee and {McGurk}, Rosalie C. and {Rich}, Jeffrey A. and {Medling}, Anne M. and {Stierwalt}, Sabrina and {Max}, Claire E. and {Evans}, Aaron S. and {Agostino}, Christopher J. and {Charmandaris}, Vassilis and {Gao}, Tianmu and {Howell}, Justin H. and {Inami}, Hanae and {Lai}, Thomas S.-Y. and {Larson}, Kirsten L. and {Martin}, Christopher D. and {Matuszewski}, Mateusz and {Mazzarella}, Joseph M. and {Neill}, James D. and {Prusinski}, Nikolaus Z. and {Remigio}, Raymond and {Sanders}, David B. and {Surace}, Jason},
        title = "{A precessing jet from an active galactic nucleus drives gas outflow from a disk galaxy}",
      journal = {Science},
         year = 2026,
        month = feb,
       volume = {391},
       number = {6788},
        pages = {911-916},
          doi = {10.1126/science.adp8989},
archivePrefix = {arXiv},
       eprint = {2601.08791},
 primaryClass = {astro-ph.GA},
       adsurl = {https://ui.adsabs.harvard.edu/abs/2026Sci...391..911K}
}

@ARTICLE{kader25,
       author = {{Kader}, Justin A. and {U}, Vivian and {Rich}, Jeffrey A. and {Bianchin}, Marina and {Linden}, Sean T. and {Medling}, Anne M. and {Diaz-Santos}, Tanio and {Privon}, George C. and {McGurk}, Rosalie and {Armus}, Lee and {Barcos-M{\~{u}}noz}, Loreto and {Canalizo}, Gabriela and {Charmandaris}, Vassilis and {Evans}, Aaron S. and {Gao}, Tianmu and {Howell}, Justin and {Inami}, Hanae and {Lai}, Thomas and {Larson}, Kirsten L. and {Malkan}, Matthew A. and {S{\'a}nchez-Garc{\'\i}a}, Mar{\'\i}a and {Martin}, Christopher D. and {Matuszewski}, Mateusz and {Max}, Claire E. and {Mazzarella}, Joseph M. and {Neill}, James D. and {Prusinski}, Nikolaus Z. and {Remigio}, Raymond and {Sanders}, David and {Song}, Yiqing and {Stierwalt}, Sabrina and {Surace}, Jason},
        title = "{Shockingly Effective: Cluster Winds as Engines of Feedback in Starburst Galaxy VV 114}",
      journal = {\apj},
         year = 2025,
        month = aug,
       volume = {988},
       number = {2},
          eid = {230},
        pages = {230},
          doi = {10.3847/1538-4357/ade38e},
archivePrefix = {arXiv},
       eprint = {2506.16624},
 primaryClass = {astro-ph.GA},
       adsurl = {https://ui.adsabs.harvard.edu/abs/2025ApJ...988..230K}
}
\bibliographystyle{aasjournal}

\end{document}